\documentclass[reprint,amsmath,amssymb]{revtex4-1}
\usepackage[utf8x]{inputenc}
\usepackage{graphicx}
\usepackage{dcolumn}
\usepackage{bm}
\usepackage{color} 
\usepackage{amsmath}
\usepackage{tensor}

\usepackage{soul} 
\usepackage[normalem]{ulem} 

\begin{document}

\preprint{APS/123-QED}


\title{Localized states and skin effect around non-Hermitian impurities in tight-binding models}

\author{Bal\'azs Het\'enyi$^{1,2}$ and Bal\'azs D\'ora$^1$}
\affiliation{$^1$Department of Theoretical Physics, Institute of Physics, Budapest University of Technology and Economics, Műegyetem rkp. 3., H-1111 Budapest, Hungary \\ and \\
 $^2$Institute for Solid State Physics and Optics, Wigner Research Centre for Physics,  H-1525 Budapest, P. O. Box 49, Hungary}

\date{\today}
\begin{abstract}
We use the generalized Bloch theorem formalism of Alase {\it et al.} [{\it Phys. Rev. Lett.} {\bf 117} 076804 (2016)] to analyze simple one-dimensional tight-binding lattice systems connected by Hermitian bonds (all with the same hopping parameter $t$), but containing one bond impurity which can be either Hermitian or non-Hermitian.   We calculate the band structure, the bulk-boundary correspondence indicator ($D_L(\epsilon)$) and analyze the eigenvalues of the lattice translation operator ($z$), for each eigenstate.  From the $z$ values the generalized Brillouin zone can be reconstructed.  If the impurity is Hermitian (and $\mathcal{PT}$-symmetric), we find a parameter regime in which two localized edge states separate from the tight-binding band.  We then simulate a non-Hermitian impurity by keeping hopping in one direction of the bond impurity the same as the rest of the tight-binding system, and varying only its reciprocal.  Again, we find a region with localized edge states, but in this case the energy eigenvalues are purely imaginary.  We also find that in this case the two zero energy eigenvectors coalesce, hence this system is an exceptional line.  We then perform an interpolative scan between the above two scenarios and find that there is an intermediate region exhibiting a non-Hermitian skin effect.  In this region a macroscopic fraction of states acquire complex energy eigenvalues and exhibit localization towards the impurity.  Our numerical results are supported by a detailed analysis of the solutions of the boundary/impurity equation.
\end{abstract}

\maketitle

\section{Introduction}

Many physical systems are best described by non-Hermitian~\cite{Bender98,Bender07} Hamiltonians: open systems~\cite{Rotter09,Malzard15,Carmichael93,Zhen15,Diehl11,Cao15,Choi10,San-Jose16,Lee14a,Lee14b}, wave systems with gain and loss~\cite{Makris08,Longhi09,Klaiman08,Regensburger12,Bittner12,Ruter10,Lin11,Feng13,Guo09,Liertzer12,Peng14,Fleury15,Chang14,Hodaei17,Hodaei14,Feng14,Gao15,Xu16,Ashida17,Kawabata17,Chen17,Ding16,Downing17}, or systems in which disorder or interaction lead to a non-Hermitian self-energy~\cite{Kozii17,Papaj19,Shen18} are some examples.    The initial motivation to study non-Hermitian systems was due to the discovery~\cite{Bender98,Bender07} that not only Hermitian matrices can produce real, and therefore physically reasonable, eigenvalues, but the set of Hermitian matrices is a subset of $\mathcal{PT}$-symmetric matrices which obey a looser set of conditions than Hermitian ones.  The breaking of $\mathcal{PT}$ symmetry, by tuning the parameters to violate this symmetry, leads to complex conjugate pairs of eigenvalues.  Non-Hermitian systems have been suggested in a number of technological applications: light-funneling~\cite{Weidemann20}, topological sensors~\cite{Budich20,McDonald18}, a topological ohmmeter~\cite{Konye24}, or unidirectional amplification~\cite{Brunelli23,Wanjura20,Xue21}.\\

The topological analysis of non-Hermitian systems~\cite{Bergholtz21,Lieu18,Yao18,Yokomizo19,Yang20} is challenging due to the fact that there are two possible localization effects if the boundaries of such a system are open: one, the topological edge states~\cite{Bernevig13,Asboth16,Qi11,Sato17} which also appear in Hermitian systems, and two, the non-Hermitian skin effect, which arises, for example, in the canonical Hatano-Nelson model~\cite{Hatano96}.  In Hermitian systems the traditional bulk boundary correspondence principle connects the existence of localized edge modes with nontrivial values of a topological invariant, which usually corresponds to a Brillouin zone integral.  For non-Hermitian systems the introduction of the generalized~\cite{Yao18,Yokomizo19,Yang20} Brillouin zone in which topological invariants can be calculated lead to results consistent with experiments.   In this work, we focus on a special class of non-Hermitian systems, lattice models which would be Hermitian, apart from a single impurity~\cite{Spring24,Sukhachov20,Molignini23,Wu23}.   A recent study~\cite{Wu23} showed that a non-Hermitian proximity effect arises when such systems gapped: this effect arises from in-gap complex energy states.  We study a more basic system, a tight-binding lattice model with Hermitian couplings between all, except one bond (non-Hermitian bond impurity).   \\

It is, of course, a question of interest whether the model studied here can be realized experimentally.   Overall, there has been great progress~\cite{Wang23} in recent years in the realization of non-Hermitian models in various experimental settings: optical waveguide arrays, photonic crystals, micro-resonator arrays, optical fiber loops, exciton-polariton systems, optomechanical systems, and ultracold atomic lattices.  The Hatano-Nelson lattice model, in particular, has been realized in coupled laser arrays~\cite{Liu22}, photonic crystals~\cite{Longhi15,Zhang25}, and audible acoustic systems~\cite{Maddi24}.  Given that the model we study is a Hermitian tight-binding lattice with a single non-Hermitian impurity, we anticipate that it can be realized in one or more of the above mentioned experimental setups.\\

We implement a generalized Bloch theorem (GBT) formalism developed by Alase et al.~\cite{Alase16,Alase17,Cobanera17} in the context of a non-Hermitian impurity problem.   Since we are dealing with tight-binding models, with no internal degrees of freedom, a simplified version of this formalism suffices.  The GBT generalizes the Bloch theorem to systems which are translationally invariant up to open of boundaries.   Since an open boundary problem can be understood as a special case of an impurity problem, the GBT easily generalizes to the single impurity problem~\cite{Teo10,Apollaro13}.  In the GBT the Hamiltonian of a given system is written explicitly in terms of lattice translation operators ($\hat{T}$).   The Hamiltonian is then separated into a bulk and a boundary part.  For a periodic system the eigenvalues of $\hat{T}$ are $|z|=1$.   The generalization consists of extending the range of $z$ to the complex plane, but requiring that both bulk and boundary Schr\"odinger equations are simultaneously satisfied.  The formalism also provides for an indicator of localization, $D_L(\epsilon)$, which diverges for localized edge or bound states.  In the works of Alase et al.~\cite{Alase16,Alase17,Cobanera17} it is used as the indicator of topological edge states via the additional requirement that $\epsilon=0$ (which is usually the result of symmetry protection).  We emphasize that the GBT has very much in common with the generalized Brillouin zone formalisms developed in the context of non-Hermitian systems~\cite{Yao18,Yokomizo19,Yang20}.  The crucial step in both cases is to diagonalize the translation operator ($\hat{T}$) whose eigenvalues ($z$) are in general complex.  For a Hermitian system with periodic boundaries, $|z|=1$, the Brillouin zone corresponds to the unit circle.  Open boundaries lead to localized edge states, usually on the real axis, while topological systems lead to edge states for which $z\in \mathbb{C}$ and $|z| \neq 1$.  Non-Hermitian systems lead to generalized Brillouin zones which encircle the origin but do not necessarily correspond to the unit circle. \\

In this paper we perform three sets of calculations using the GBT.   We present the energy spectra, the bulk-boundary correspondence indicator, $D_L(\epsilon)$ and an analysis of the eigenvalues ($z$) of the lattice translation operator, $\hat{T}$.  Our first calculation shows that if the impurity is Hermitian, there exists a region with localized edge states, whose energy eigenvalues separate from the tight-binding band, have $z$ values off the Brillouin zone on the real axis, and which exhibit a diverging $D_L(\epsilon)$.   All other states are extended Bloch type states, for which $|z|=1$.  In the second example calculation, we make the impurity non-Hermitian by keeping the hopping parameter in one direction equal to the rest of the tight-binding system, but varying its reciprocal bond.  We find a region in which, again, there exists a pair of localized edge states, exhibiting a diverging $D_L(\epsilon)$, but these states have $z$ values off the Brillouin zone on the imaginary axis, rather than the real axis.  The imaginary energy eigenvalues are negatives of each other.  We also perform an interpolative scan between these two situations and find an intermediate region in which none of the states have finite $D_L(\epsilon)$ values.  However, the energy values become complex for a fraction of the states, simultaneously, the $z$ values for these states move off the unit circle in reciprocal pairs.   Finite size scaling indicates that this fraction becomes constant in the thermodynamic limit, implying a macroscopic fraction.  The states exhibit a skin effect, localization towards the impurity.   The other states remain on the unit circle, and are extended states.   Our numerical results are all strongly supported by the solutions of the boundary/impurity equation (BIE). \\

The curious reader is in for the following ride.  In the following section we present the necessary background of the GBT method.  In section \ref{sec:TBImp} we present the models we study and give the expression of the bulk-boundary correspondence indicator in the tight-binding, single impurity context.  In section \ref{sec:bie} we derive the BIE.  In section \ref{sec:rslts} the numerical results for the three different bond impurity types are presented and analyzed.  Before concluding (section \ref{sec:cnclsn}) we show that the BIE is in agreement with all our numerical results..

\section{Background: the bulk boundary separation method}

\label{sec:bckgrnd}

In this section we introduce a simplified version of the bulk-boundary separation (GBT) formalism developed for topological systems with edge states by Alase et al. \cite{Alase16,Alase17,Cobanera17}.   Our simplifications make the formalism appropriate for simple tight-binding models without internal degrees of freedom.  For the full details of the method, see Refs. \cite{Alase16,Alase17,Cobanera17}.\\

Given a system of size $L$ with sites $j=0,...,L-1$.   We write the Hamiltonian, ordered in a particular way,  as,
\begin{equation}
\label{eqn:H1}
\hat{H} = \frac{1}{2} \sum_{r=1}^R \left( \sum_{j=0}^{L-1-r} a_j^\dagger h_r a_{j+r} + \sum_{j=L-r}^{L-1} a_j^\dagger g_r a_{j+r-L} + \mbox{H.c.} \right),
\end{equation}
where $h_r$ and $g_r$ denote the hopping parameters in the bulk and at the boundary, respectively.  $R$ denotes the range of the Hamiltonian, the largest distance between unit cells connected by hopping.   The eigenvalue equation of $\hat{H}$ is
\begin{equation}
\hat{H} | \epsilon \rangle = \epsilon | \epsilon \rangle.
\end{equation}
We seek to solve this eigenvalue equation, but taking advantage of the ordering of terms in Eq. (\ref{eqn:H1}).  We will construct the eigenstates and construct the eigenfunctions using translation operators appropriate to the given boundary conditions.

Periodic boundary conditions correspond to $g_r = h_r$, while $g_r = 0$ occurs when the boundaries are open $\forall r$.   Intermediate values of $g_r$ can be interpreted as an impurity in a system with periodic boundary conditions.  Notice that the index $r$ in both $h_r$ and $g_r$ refers to the distance between sites.   We can now introduce the basis ${ | j \rangle }$,
\begin{equation} 
| j \rangle  = a^\dagger_{j}|0\rangle.
\end{equation}
and write the Hamiltonian using shift operators, 
\begin{equation}
\label{eqn:H2}
\hat{H} = \sum_{r=0}^R [ h_r  \hat{T}^r   + g_r  (\hat{T}^\dagger)^{L-r} + \mbox{H. c.}],
\end{equation}
where the left-shift operator satisfies $\hat{T}|j\rangle = |j-1\rangle, \forall j \neq 0$, and $\hat{T}|0\rangle = 0$.  $\hat{T}^\dagger$ implements the corresponding right-shift.   $\hat{T}^r$ denotes multiplying the matrix $\hat{T}$ (regular matrix multiplication) $r$ times.\\

Periodic boundary conditions can be implemented by using the operators $\hat{V} = \hat{T} + (\hat{T}^\dagger)^{L-1}$ (and letting $g_r = h_r$).  The Hamiltonian now becomes
\begin{equation}
\hat{H} = \sum_{r=0}^R [ h_r  \hat{V}^r  +  \mbox{H. c.}],
\end{equation}
and $\hat{H}, \hat{V}$ and $\hat{V}^\dagger$ from a commuting set.  Using the generalized $z$-transformed lattice basis, 
\begin{equation}
\label{eqn:z}
| z \rangle = \frac{1}{\sqrt{N(z)}} \sum_{j=1}^{L} z^j | j \rangle, z \in \mathbb{C}, z \neq 0.
\end{equation}
one can now obtain the "reduced bulk Hamiltonian", which in this case, due to no internal degrees of freedom, is a one-by-one matrix, $h_B(z)$ equal to the energy eigenvalues $\epsilon$,
\begin{equation}
\label{eqn:h_B}
\epsilon = h_B(z) = \sum_{r=0}^R (z^r h_r + z^{-r} h_r^{-1}).
\end{equation}
This equation relates the complex number $z$ to the energy eigenvalues $\epsilon$.  In a periodic system $z=e^{ik}$ is on the unit circle ($k$ is the crystal momentum).  The first Brillouin zone is defined as $k = 2 \pi q/L$, $q = 0,...,L-1$.  The eigenvectors of $\hat{H}$, which are ordinary Bloch states, take the form $|\epsilon \rangle = |z \rangle$ with eigenvalue of $\epsilon = h_B(z) = -2t\cos(k)$.   The $z$-transform in this case is a discrete Fourier transform.\\

For systems with open boundary conditions, where lattice translation symmetry is broken at the boundaries, the Fourier transform fails to diagonalize $\hat{H}$, and the operators $\hat{T}$ and $\hat{T}^\dagger$ do not share a common eigenbasis.  The diagonalization procedure of GBT first extends the range of $z$ to the entire complex plane, allowing for decaying edge states at the boundaries.  Second, they split the eigenvalue equation into two pieces, a bulk and a boundary equation, using two projector operators, $\hat{P}_B = \sum_{j=R}^{L-1-R} | j \rangle \langle j |$, the bulk projector, and $\hat{P}_\partial = \mathbb{I}_L - \hat{P}_B$, where $\mathbb{I}_L$ denotes the identity in the Hilbert subspace associated with lattice sites.   The bulk and boundary equations read, $\hat{P}_B \hat{H} | \epsilon \rangle = \epsilon \hat{P}_B |\epsilon \rangle$ and  $\hat{P}_\partial \hat{H} | \epsilon \rangle = \epsilon \hat{P}_\partial |\epsilon \rangle$, respectively.  As a result of this separation, we obtain simultaneous relative eigenvectors of the bulk-projected $\hat{T}$ and $\hat{T}^\dagger$ operators.  \\

Putting the above into practice, one first solves the reduced bulk equation, which is a characteristic polynomial equation of $h_B(z)$,
\begin{equation}
\label{eqn:characteristic}
P(\epsilon,z) = z^R|h_B(z) - \epsilon | = 0,
\end{equation}
There are $2R$ values of $z_l$ which satisfy Eq. (\ref{eqn:characteristic}) for a given $\epsilon$ ($l=1,...,2R$).  The full solution for a given eigenvalue $\epsilon$ can be written,
\begin{equation}
|\epsilon \rangle = \sum_{l=1}^{2R}  \alpha_l | z_l(\epsilon) \rangle, \alpha_l \in  \mathbb{C}.
\end{equation}
The coefficients $\alpha_l$ are determined from solving the boundary equation, $\hat{P}_\partial (\hat{H} - \epsilon \mathbb{I})|\epsilon \rangle = 0$.   One can derive an equation for $\alpha_l$ by first using the $2R$ boundary states $\{|j\rangle, 0\leq j \leq R-1, L-R \leq j \leq L-1 \}$.  We can then form the $2R$ equations,
\begin{equation}
\sum_{l=1}^{2R}\langle j | \hat{P}_\partial (\hat{H} - \epsilon \mathbb{I})  | z_l(\epsilon) \rangle \alpha_{l} = 0.
\end{equation}
To construct the bulk-boundary indicator, one first forms a matrix,
\begin{equation}
[B_L(\epsilon)]_{jl} = \langle j | \hat{P}_\partial (\hat{H} - \epsilon \mathbb{I})  | z_l(\epsilon) \rangle.
\end{equation}
The bulk-boundary indicator defined in Ref. \cite{Alase16} reads as,
\begin{equation}
\label{eqn:D}
D_L(\epsilon) = \log \det \{ B^\dagger_L(\epsilon) B_L(\epsilon) \}.
\end{equation}
Alase et al. derived this quantity and suggested that $D_\infty(0)$ is the equivalent of the topological invariant for systems with open boundary conditions.  In topological systems symmetry protection guarantees that edge localized states are $\epsilon=0$ modes.  We will calculate $D_L(\epsilon)$ for entire bands for the models studied here.\\

\begin{figure}[t]
 \centering
 \includegraphics[width=8cm,keepaspectratio=true]{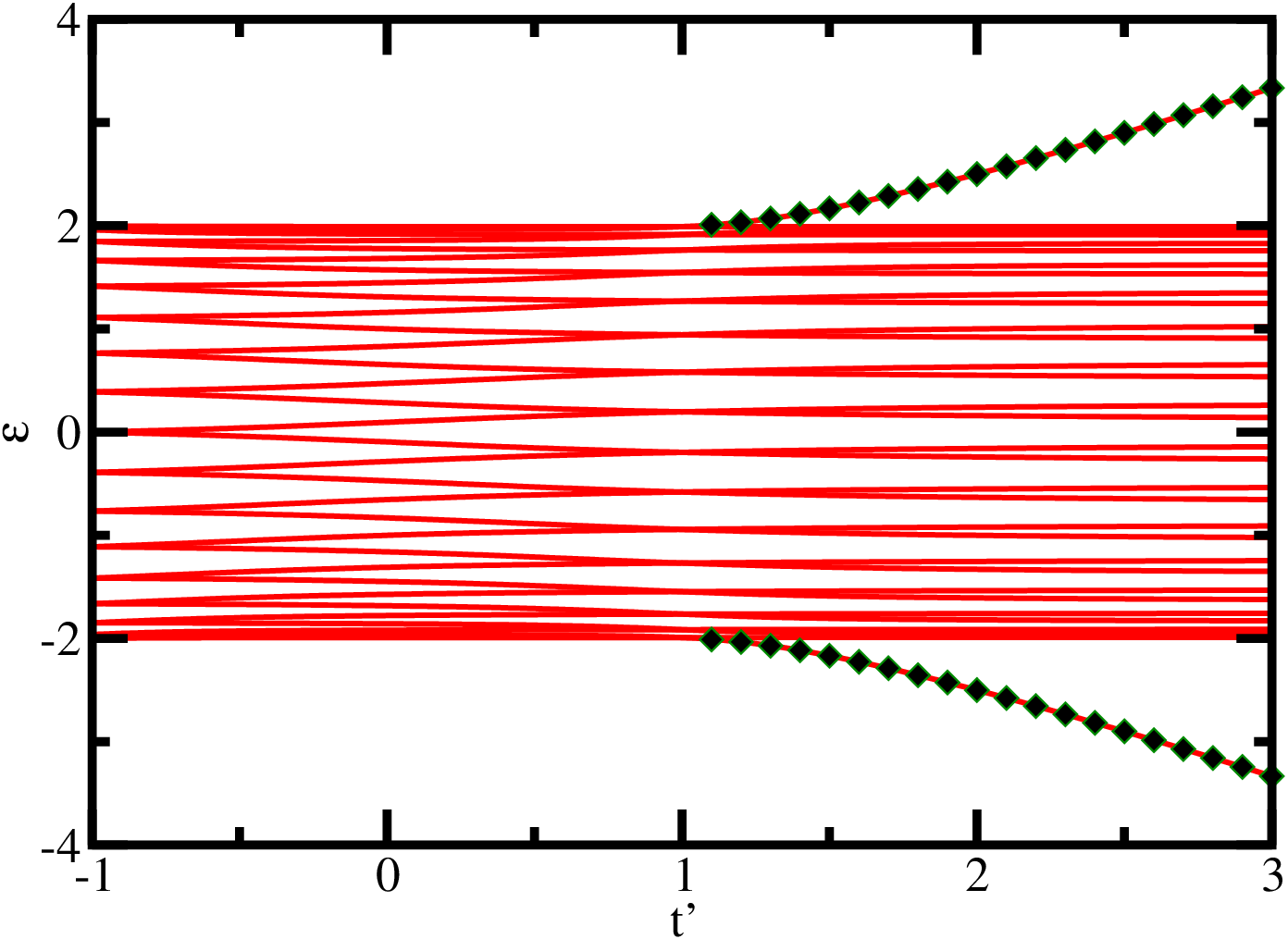}
 \caption{Band structure of a tight-binding model with a Hermitian bond impurity for a system of $L=32$ as a function of $t'$.   As $t'=1$ two states, one from the bottom, one from the top of the band, separate from the overall band.  We calculated the bulk-boundary correspondence indicator ($D_L(\epsilon)$) and found it to be divergent for these states, but not for any of the others.}
  \label{fig:HBondImp_ev}
\end{figure}

\begin{figure}[t]
 \centering
 \includegraphics[width=8cm,keepaspectratio=true]{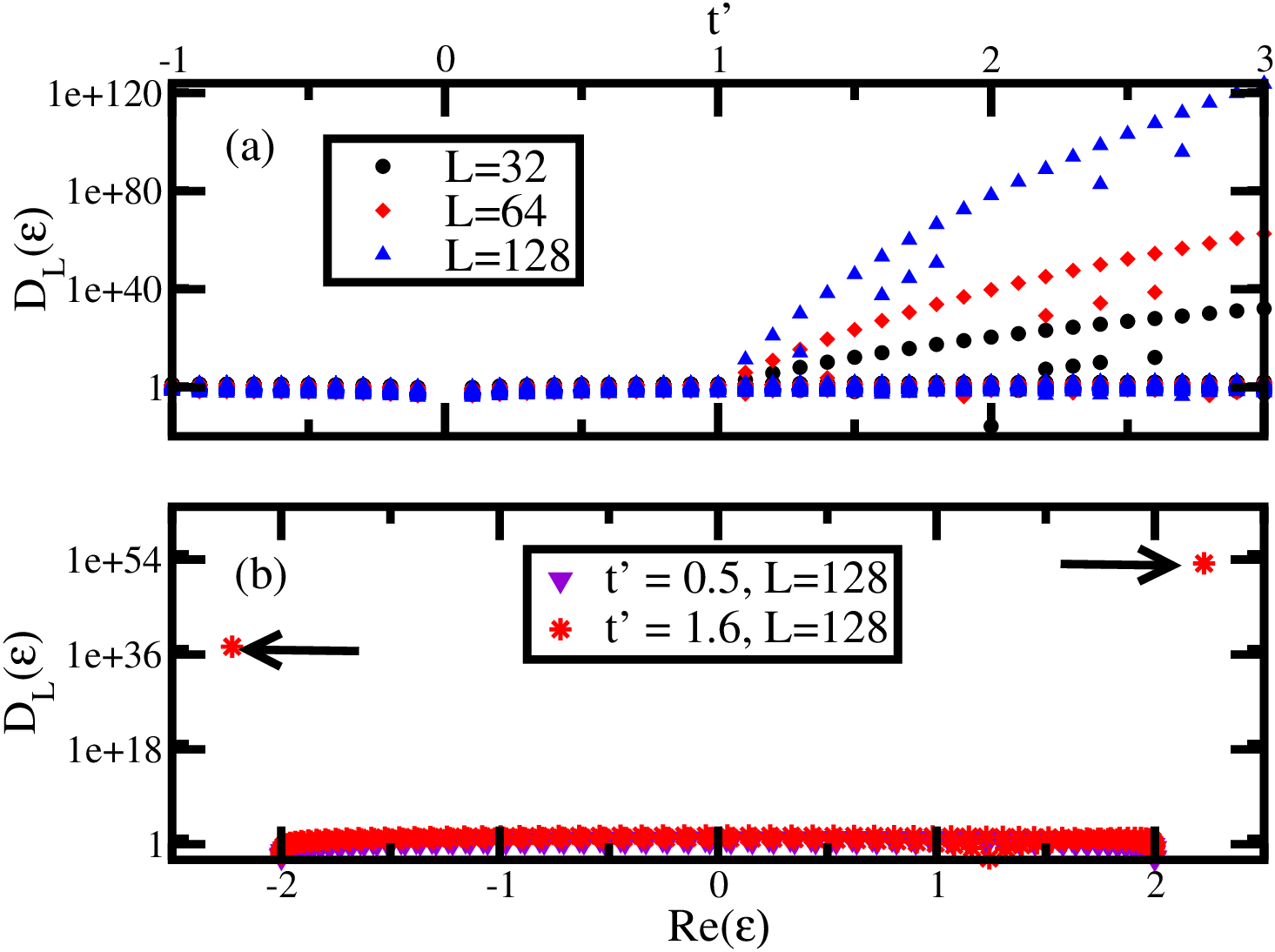}
 \caption{Bulk-boundary correspondence indicator, $D_L(\epsilon)$ for a Hermitian system with a Hermitian bond impurity.   Panel (a) shows $D_L(\epsilon)$ as a function of $t'$ for all states for three system sizes.  As $t'$ increases above $t'=1$ for a certain set of states $D_L(\epsilon)$ diverges.  The larger the system size, the steeper the increase.  Panel (b) shows $D_L(\epsilon)$ as a function of the energy eigenvalue, for two values of $t'$, $t'=0.5,1.6$.  The arrows indicate the extremal states which separated from the band and are localized for $t'=1.6$.}
 \label{fig:HBondImp_D}
\end{figure}

\begin{figure}[t]
 \centering
 \includegraphics[width=8cm,keepaspectratio=true]{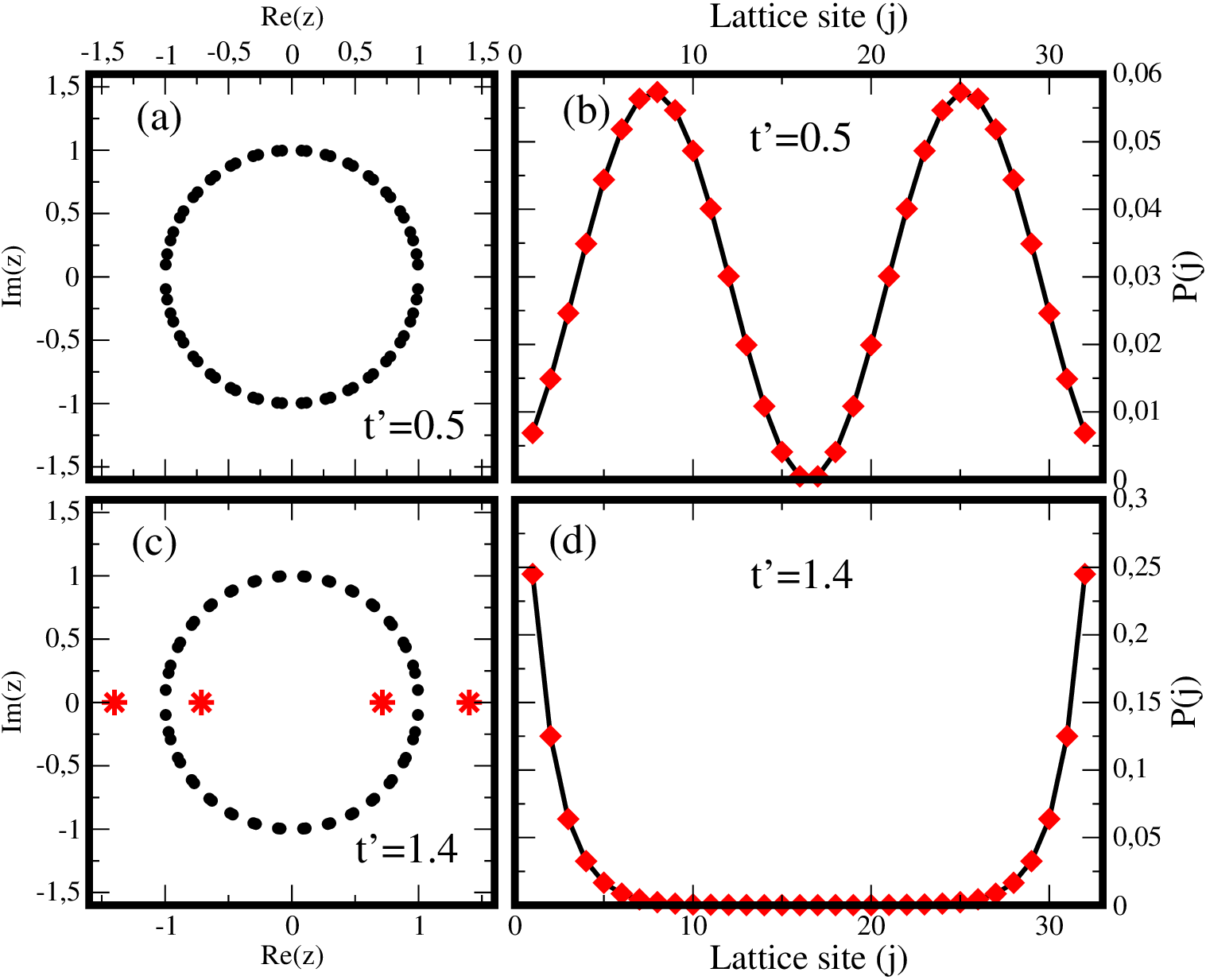}
 \caption{Tight-binding model with a Hermitian bond impurity ($L=32$), panel (a) $z$-values for $t'=0.5$, all values fall on the Brillouin zone, (b) probability distribution for the state indicated in dotted lines in Fig. \ref{fig:HBondImp_ev}, (c) $z$-values for $t'=1.4$, red asterisks indicate $z$ values for states which separated from the band in Fig. \ref{fig:HBondImp_ev}, (d) probability distribution for one of the separated states.}
 \label{fig:HBondImp_z}
\end{figure}

\section{Tight-binding model with a bond impurity}

\label{sec:TBImp}

As a first example, we study the simple tight binding model with periodic boundary conditions, but one bond modified into a bond impurity, which can be taken to be Hermitian or non-Hermitian (Hatano-Nelson).  Such a Hamiltonian reads,
\begin{equation}
\label{eqn:HNBondImp_H}
H = -t \sum_{j=1}^{L-1} (c_j^\dagger c_{j+1} + c_{j+1}^\dagger c_j ) - t'' c_L^\dagger c_1 - t' c_1^\dagger c_L,
\end{equation}
where $t$ denote the hopping integral on all bonds, except the bond connecting site $1$ and $L$, where the hopping integrals are not necessarily equal $t'$ and $t''$.   \\

In the Hermitian case ($t'=t''$), the model is $\mathcal{PT}$-symmetric.  Defining the parity ($\mathcal{P}$) and time-reversal ($\mathcal{T}$)as
\begin{equation}
\mathcal{P} c_j \mathcal{P}^{-1} = c_{L+1-j},\hspace{.1cm}\mathcal{T} = \hat{K},
\end{equation}
where $\hat{K}$ denotes complex conjugation, it can be shown that
\begin{equation}
[H,\mathcal{P}\mathcal{T}]_- = 0,
\end{equation}
where $[,]_-$ denotes the commutator.  In parameter ranges considered here ($t$, $t'$, and $t''$ all real), the model is anti-$\mathcal{CP}$-symmetric.  One can define the charge conjugation operator,
\begin{equation}
\mathcal{C}c_j \mathcal{C}^{-1} = c_j^\dagger,\hspace{.1cm} \mathcal{C} c_j^\dagger \mathcal{C}^{-1} = c_j.
\end{equation}
One can then show that,
\begin{equation}
[H,\mathcal{CP}]_+ = 0,
\end{equation}
where $[,]_+$ denotes the anticommutator.\\

It is of interest to consider the effect of the symmetry operator on the basis states and possible state furnished by the GBT formalism.  In particular, for extended states ($z=e^{ik}$) it holds that,
\begin{equation}
\mathcal{P T} | z \rangle = | z \rangle,
\end{equation}
whereas, for states localized near the impurity, with $z$ values purely on the real axis, ($z=e^\kappa$),
\begin{equation}
\mathcal{P T} | z \rangle = | z^{-1} \rangle.
\end{equation}
More general $|z\rangle$ states do not satisfy $\mathcal{PT}$-symmetry.  For the anti-$\mathcal{CP}$-symmetric case, for any $z$, it holds that,
\begin{equation}
\mathcal{CP}| z \rangle = |z^{-1}\rangle.
\end{equation}

For this Hamiltonian, $R=1$, $h_0=0$, and $h_1 = -t$,  leading to,
\begin{equation}
h_B(z) = -t (z + z^{-1}).
\end{equation}
In our calculations, at particular values of $t, t'$ and $t''$, for a given system size, we diagonalize the Hamiltonian, giving the energy eigenvalues and eigenvectors.   For a given eigenvalue, $\epsilon$, we find the $z$ values (two for each energy eigenvalue) from $h_B(z_i) = \epsilon$, with $i=1,2$.   This means that the matrix,  $B^\dagger_L(\epsilon_j) B_L(\epsilon_j)$ is a two-by-two matrix.  Applying the steps of section \ref{sec:bckgrnd} results in the following bulk-boundary correspondence indicator,
\begin{equation}
D_L(\epsilon) = \log \mbox{det} M
\end{equation}
where
\begin{equation}
M_{ij} = \left[ \frac{1}{\sqrt{N(z_i) N(z_j)}}\left( (f^{(1)}_i)^* f^{(1)}_j + (f^{(L)}_i)^* f^{(L)}_j\right)\right],
\end{equation}
and where $N(z_i)$ is the normalization constant for $| z_i \rangle$ (see Eq. (\ref{eqn:z})), and 
\begin{eqnarray}
f^{(1)}_i &=& -t z_i^2 - t' z_i^L - \epsilon_j z_i \\ \nonumber
f^{(L)}_i &=& -t z_i^{L-1} - t'' z_i - \epsilon_j z_i^L. 
\end{eqnarray}

\begin{figure}[t]
 \centering
 \includegraphics[width=8cm,keepaspectratio=true]{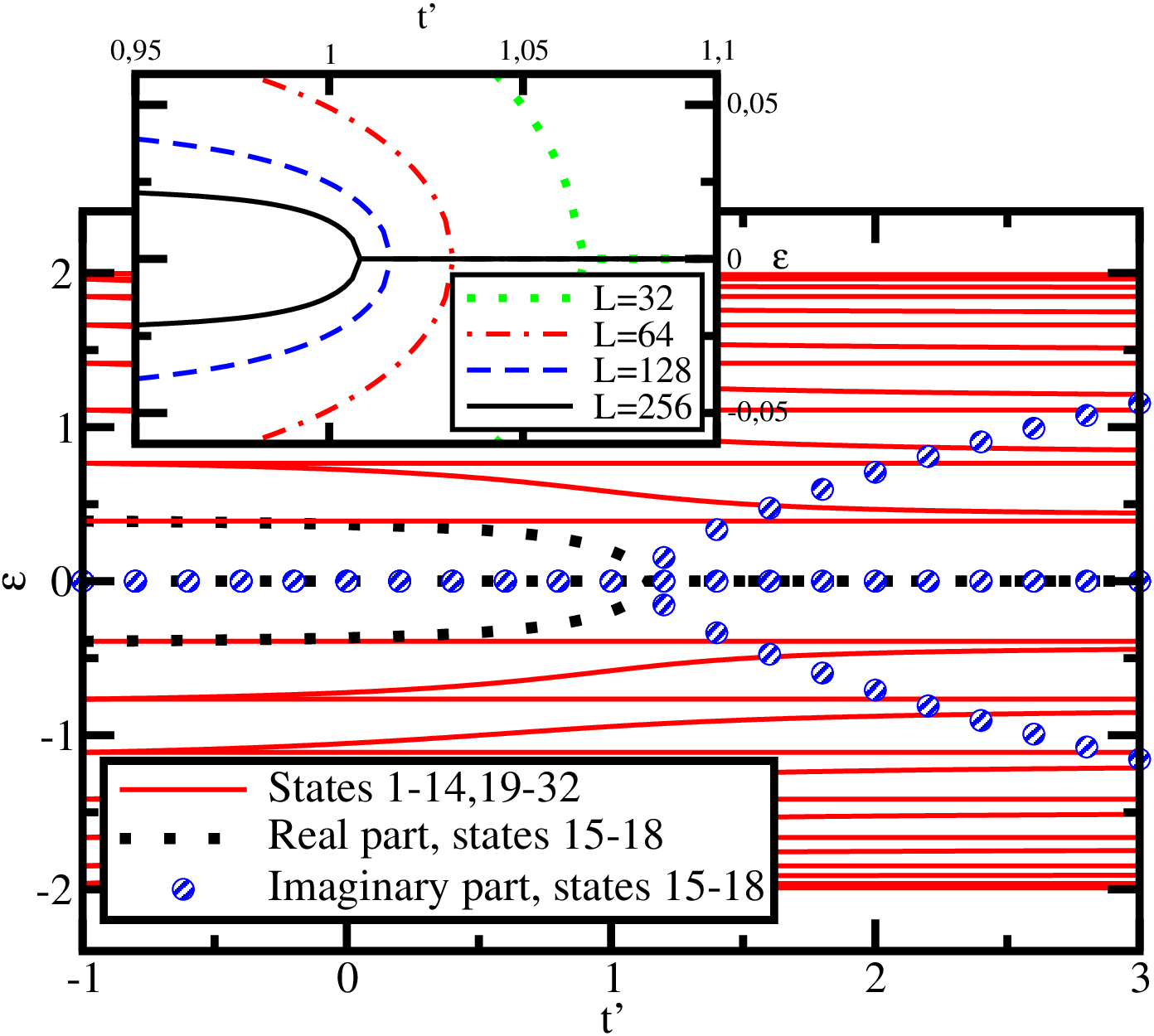}
 \caption{Band structure of a tight-binding model with a Hatano-Nelson impurity (Eq. (\ref{eqn:HNBondImp_H})) for a system of $L=32$ as a function of $t'$.  The states are ranked in order of their real parts.  The full thin red lines indicate energy eigenvalues of states which remain real throughout (rank $1-14$ and $19-32$).  The real part of states $15,16,17,18$ are indicated with black dashed lines, while their imaginary parts are blue spheres with striped filling.   These states are close to the middle of the band, the $16$th and $17$th ones have real parts of zero throughout.  States $15$ and $18$ have finite real parts at $t'=-1$, which go to zero at $t'=1.13...$, where the imaginary parts start to branch out from zero.  Upon increase in system size the $t'$ at which this occurs approaches $t'=1$, as shown in the inset.}
 \label{fig:HNBondImp_ev}
\end{figure}

\begin{figure}[t]
 \centering
 \includegraphics[width=8cm,keepaspectratio=true]{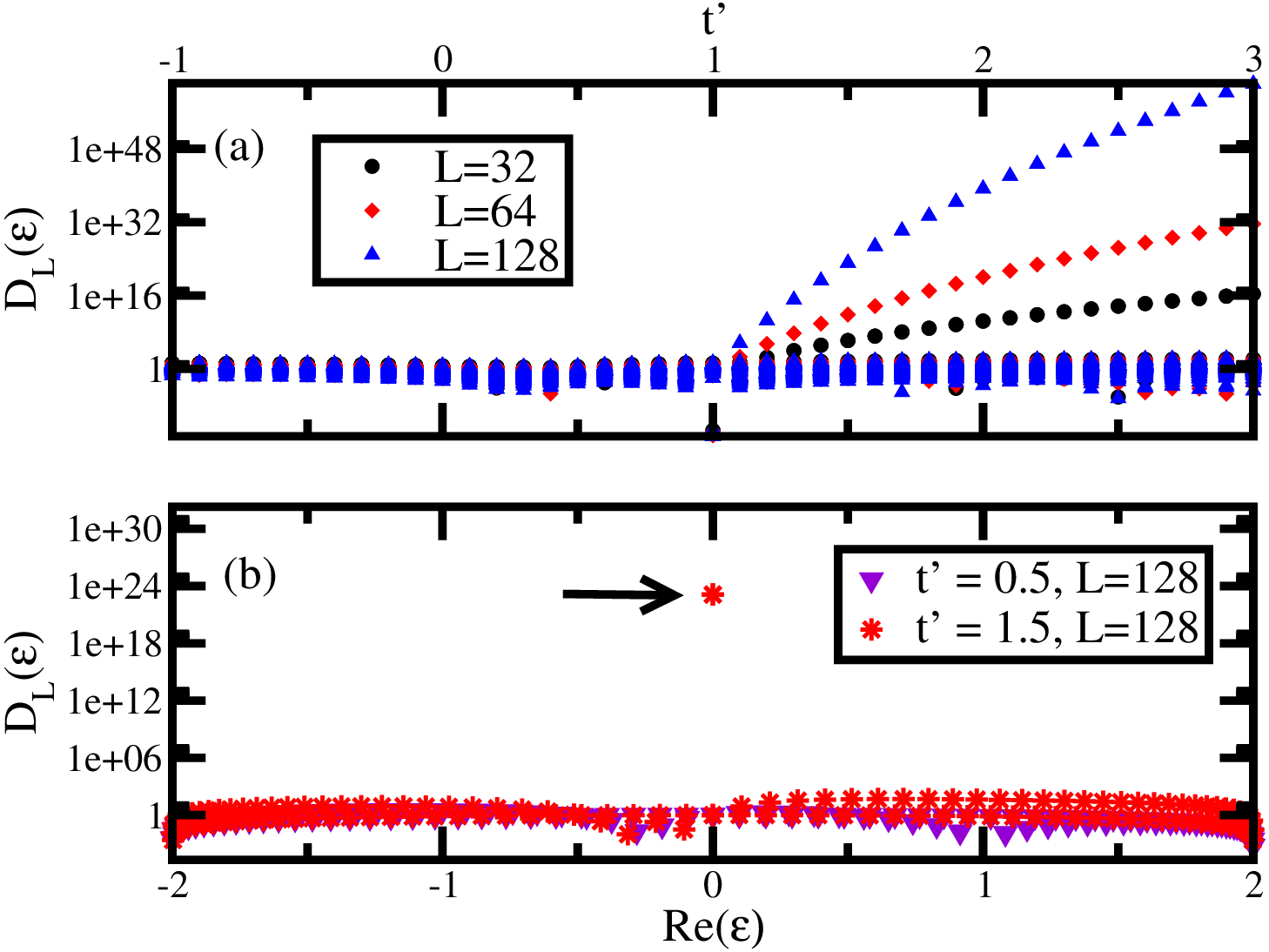}
 \caption{Bulk-boundary correspondence indicator, $D_L(\epsilon)$ for a system in which $t''=t$ as a function of $t'$.   Panel (a) shows $D_L(\epsilon)$ as a function of $t'$ for all states for three system sizes ($L=32, 64, 128$).  As $t'$ increases above $t'=1$ one set of results for each system size starts to increase by orders of magnitude (the $y$ axis is on a logarithmic scale).  The larger the system size, the steeper the increase.  Panel (b) shows $D_L(\epsilon)$ as a function of the real part of the energy eigenvalue, for two values of $t'$, $t'=0.5,1.5$.  The arrow indicates two data points (which fall in the same point on the figure) corresponding to the two states with zero real part and finite imaginary parts of opposite sign.  $D_L(\epsilon)$ for these states is many orders of magnitude larger than for all other states.}
 \label{fig:HNBondImp_D}
\end{figure}

\section{The boundary/impurity equation}

\label{sec:bie}

In this section we derive what we will call the boundary/impurity equation (BIE).  The BIE arises when one acts with a Hamiltonian (which can be PBC, OBC, or one including an impurity) on a state of the form $|z\rangle$ (of the form in Eq. (\ref{eqn:z})).  In PBC, $|z\rangle$ is an eigenstate.  In the case of OBC and the impurity problem, $|z\rangle$ is not an eigenstate, because extra terms are generated due to the open boundary or due to the impurity.  One can make these terms disappear by taking a linear combination of two $z$-states, in our case, $|z\rangle$ and $|z^{-1}\rangle$, and one can then make the boundary/impurity terms cancel.   For a simple tight-binding model with OBC this derivation was carried out by Alase et al.~\cite{Alase16}.   We extend this derivation to the case of a single impurity in a PBC system.   In section \ref{sec:nlss} we will use the boundary/impurity equation to support our numerical results of Section \ref{sec:rslts}.\\

We follow a derivation in Ref. \cite{Alase16}, but adapt it to the model with a single impurity.  Our starting point is the Hamiltonian in Eq. (\ref{eqn:HNBondImp_H}), but here we write it in bra-ket notation for convenience, as 
\begin{equation}
\label{eqnA:H}
H = -t \sum_{j=1}^{L-1} (|j\rangle \langle j+1 | + |j+1 \rangle \langle j |) - t'' |L \rangle \langle 1 |  - t' |1 \rangle \langle  L |,
\end{equation}
We consider the action of $H$ on a state of the form,
\begin{equation}
| z \rangle = \sum_{j=1}^L z^j |j \rangle.
\end{equation}
The action of $H$ on the state $|z\rangle$ can be written as,
\begin{equation}
H | z \rangle = \epsilon |z \rangle + (t z^{L+1} - t'' z) |L\rangle  + (t - t' z^L)|1\rangle.  
\end{equation}
where $\epsilon = -t(z+z^{-1})$.  The state $|z\rangle$ by itself is not an eigenstate of $H$, due to the boundary terms appearing in the previous equation.  We can explicitly construct an eigenstate by taking the linear combination,
\begin{equation}
| \epsilon \rangle = \alpha |z \rangle + \beta | z^{-1} \rangle.
\end{equation}
In this case,
\begin{eqnarray}
H | \epsilon \rangle &=& \epsilon | \epsilon \rangle + \alpha [(t z^{L+1} - t'' z) |L\rangle  + (t - t' z^L)|1\rangle] \hspace{.3cm}\\
& & + \beta  [(t z^{-L-1} - t'' z^{-1}) |L\rangle  + (t - t' z^{-L})|1\rangle]. \nonumber
\end{eqnarray}
One can choose the parameters $\alpha$ and $\beta$ so that $| \epsilon \rangle$ becomes an eigenstate with eigenvalues $\epsilon$.  To achieve this one can set the boundary term to zero,
\begin{eqnarray}
0 &=& + \alpha [(t z^{L+1} - t'' z) |L\rangle  + (t - t' z^L)|1\rangle] \hspace{.3cm}\\
& & + \beta  [(t z^{-L-1} - t'' z^{-1}) |L\rangle  + (t - t' z^{-L})|1\rangle]. \nonumber
\end{eqnarray}
By taking the scalar product with the bra states $\langle 1 |$ and $\langle L |$ we obtain the two-by-two matrix equations,
\begin{equation}
\begin{pmatrix} t z^{L+1} - t'' z & t z^{-L-1} - t'' z^{-1} \\  t - t' z^L & t - t' z^{-L} \end{pmatrix} \begin{pmatrix} \alpha \\ \beta \end{pmatrix} = 0
\end{equation}
Taking the determinant of the two-by-two matrix and setting it to zero results in the boundary/impurity equation (BIE),
\begin{equation}
\label{eqn:bie}
z^2(tz^L-t'')(tz^L-t') - (t-t''z^L)(t-t'z^L) = 0.
\end{equation}
One can determine the spectrum of the Hamiltonian by solving this equation for all possible values of $z$ and using the relation $\epsilon = -t(z + z^{-1})$. \\

To lend credence to the BIE, we consider two simple cases.  For a tight-binding model with PBC, $t'=t''=t$, and the $z$ values satisfy,
\begin{equation}
z = e^{i k_m}, \hspace{.1cm} k_m = \frac{2 \pi m}{L}, \hspace{.1cm} m = 0,...,L-1, 
\end{equation}
which are the basis functions of the usual solutions of the model.  If OBC are applied, then $t'=t''=0$, and in this case,
\begin{equation}
z = e^{i k_m}, \hspace{.1cm} k_m = \frac{ \pi m}{L+1}, \hspace{.1cm} m \in \mathbb{Z},
\end{equation}
which are the basis functions for the tight-binding analog of "particle-in-a-box" states.  \\

\section{Results}

\label{sec:rslts}

In all our calculations we set the energy scale to be $t$, the hopping strength of the Hermitian bulk system.

\subsection{Hermitian bond-impurity, $t'=t''\neq t$}

\label{ssec:r1}

Fig. \ref{fig:HBondImp_ev} shows the energy spectrum of an $L=32$ size system with a Hermitian bond impurity as a function of $t'$.   For the region $t'<1$  the states are clustered, forming a band.  At $t'=1$ a qualitative change occurs, two states separate from the band, one goes increases ($\epsilon < -2$), one decreases ($\epsilon > 2$).  The black diamonds indicate that for these states the bulk-boundary correspondence indicator, $D_L(\epsilon)$ diverges as a function of system size, as shown in Fig. \ref{fig:HBondImp_D}.  Fig. \ref{fig:HBondImp_D}(a) shows $D_L(\epsilon)$ for three different system sizes as a function of $t'$.   The evolution of all the states are shown.  The $y$-axis is logarithmic.  Clearly, for $t'>1$, $D_L(\epsilon)$ increases for some states, and this increase is steeper when the system size is increased.  We interpret it as numerical evidence for a divergence.  Fig. \ref{fig:HBondImp_D}(b) shows two "cross-sections" of Fig. \ref{fig:HBondImp_D}(a), in other words, $D_L(\epsilon)$ as a function $\epsilon$, the energy eigenvalue, for two selected values of $t'$.  For $t'=0.5$ no diverging $D_L(\epsilon)$ are found, but for $t'=1.6$, the two outermost states (one at $\epsilon<-2$, one at $\epsilon>2$) exhibit $D_L(\epsilon)$ values which are many orders of magnitudes larger than all the others. \\

Fig. \ref{fig:HBondImp_z} shows the calculated $z$ values and the probability density associated with the same wavefunction (lowest energy state for $t'=1.4$, second lowest energy state for $t'=0.5$, because the lowest and second lowest state cross at $t'=1$).  For $t'=0.5$ the $z$ values all fall on the Brillouin zone ($|z|=1$), but the edge states for $t'=1.4$ give four $z$ values which are not on the unit circle, but come in reciprocal pairs on the real axis.  The wavefunction is delocalized for $t'=0.5$, but strongly localized near the Hermitian impurity for $t'=1.4$.

\begin{figure}[t]
 \centering
 \includegraphics[width=8cm,keepaspectratio=true]{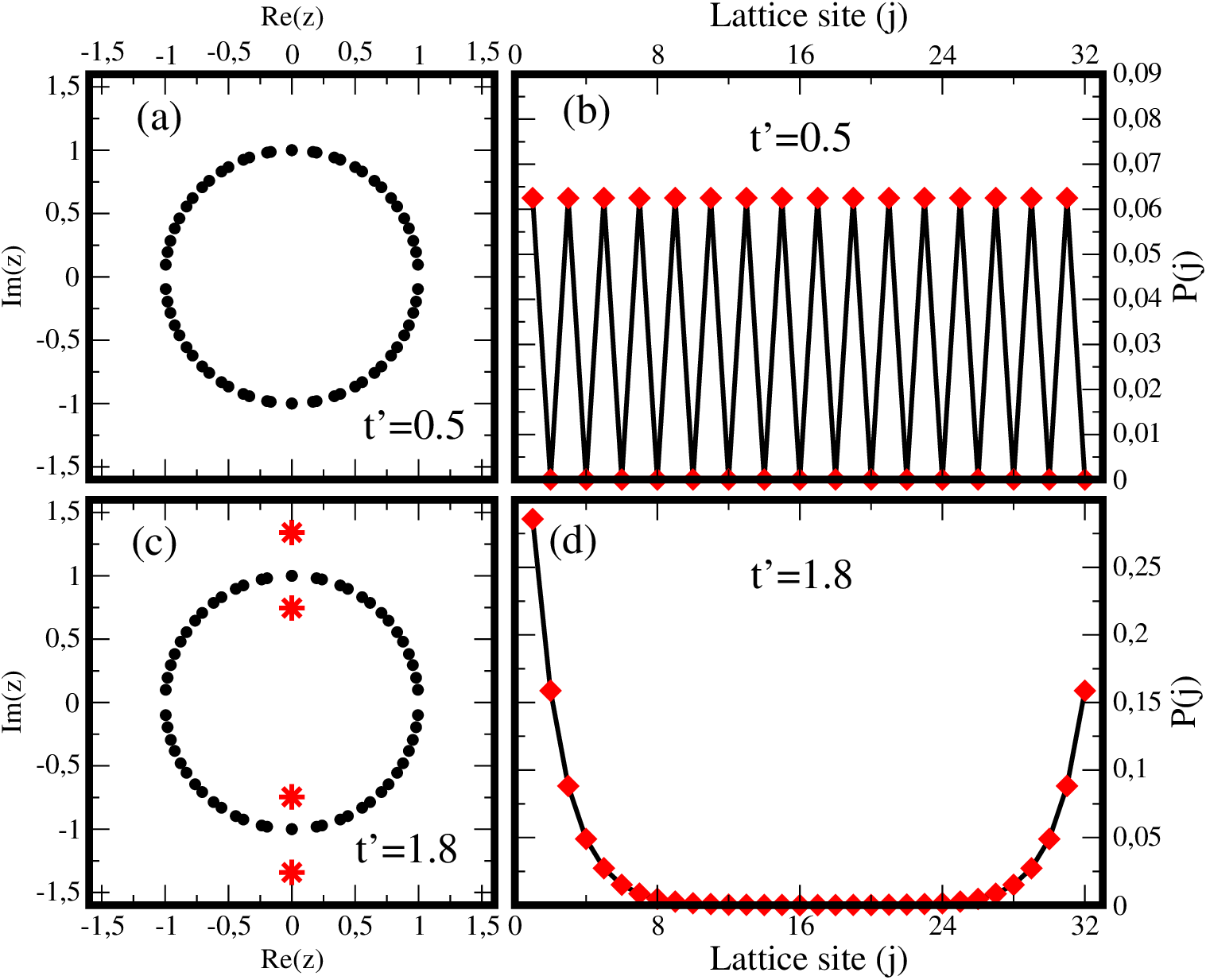}
 \caption{Tight-binding model with a Hatano-Nelson impurity ($L=32$), panel (a) $z$-values for $t'=0.5$, all values fall on the Brillouin zone, (b) probability distribution for a state with zero energy eigenvalue (both real and imaginary parts), (c) $z$-values for $t'=1.8$, red asterisks indicate $z$ values for states with zero real part but finite imaginary part of the energy eigenvalue, (d) probability distribution for the state with zero real part of the energy eigenvalue and finite imaginary part.}
 \label{fig:HNBondImp_z}
\end{figure}

\subsection{Non-Hermitian bond impurity, $t=t''\neq t'$}

\label{ssec:r2}

The band structure, for a system of $L=32$, is shown in Fig. \ref{fig:HNBondImp_ev}, as a function of $t'$ ranked in order of the real parts of the eigenvalues.    The states indicated by thin solid red lines form a large part of the band structure, these states do not exhibit edge localization.  In the middle of the band, there are four states of interest.  For any $t'$  two of these states are always zero (both real and imaginary parts).   The other two  have finite real parts for the region $t'<1$, one larger than zero, one smaller (indicated with dashed black lines).  For the two  zero eigenvalue states the eigenvectors coalesce, hence the system itself is an exceptional line.   The imaginary parts in this region of these two states are zero.  This changes at $t' \approx 1$ at a slightly larger value for the finite system shown, but the inset shows the real parts of the two states in question converging closer to $t'=1$ as the system size is increased.  For the region $t'>1$, these two states acquire finite imaginary parts, symmetrically around $\epsilon=0$.  We have done studies with larger system sizes.  What changes is that there are more states which are not participating in edge localization, the ones that do are still only the two states near $\epsilon=0$.\\
\begin{figure}[t]
 \centering
 \includegraphics[width=8cm,keepaspectratio=true]{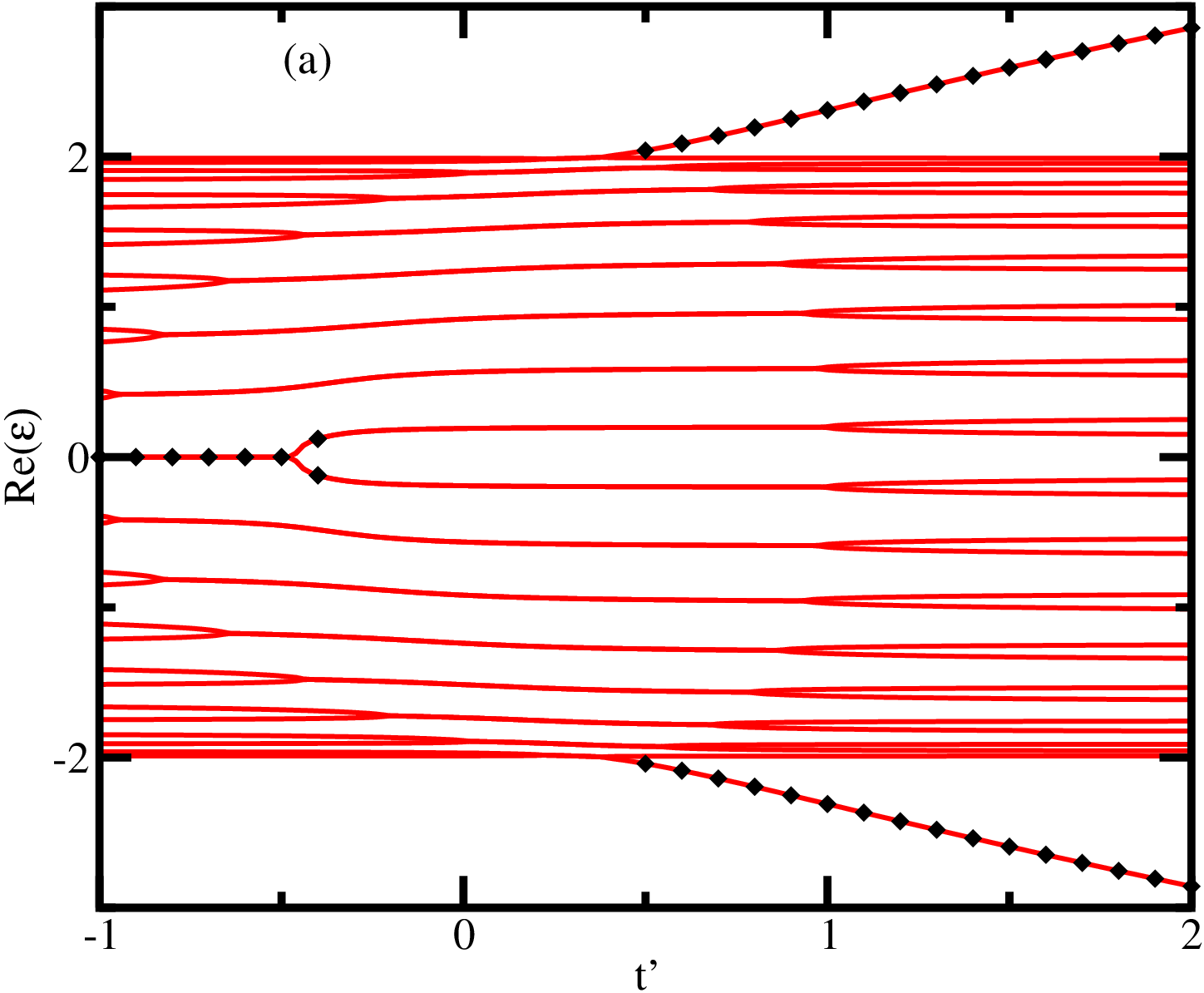}
 \includegraphics[width=8cm,keepaspectratio=true]{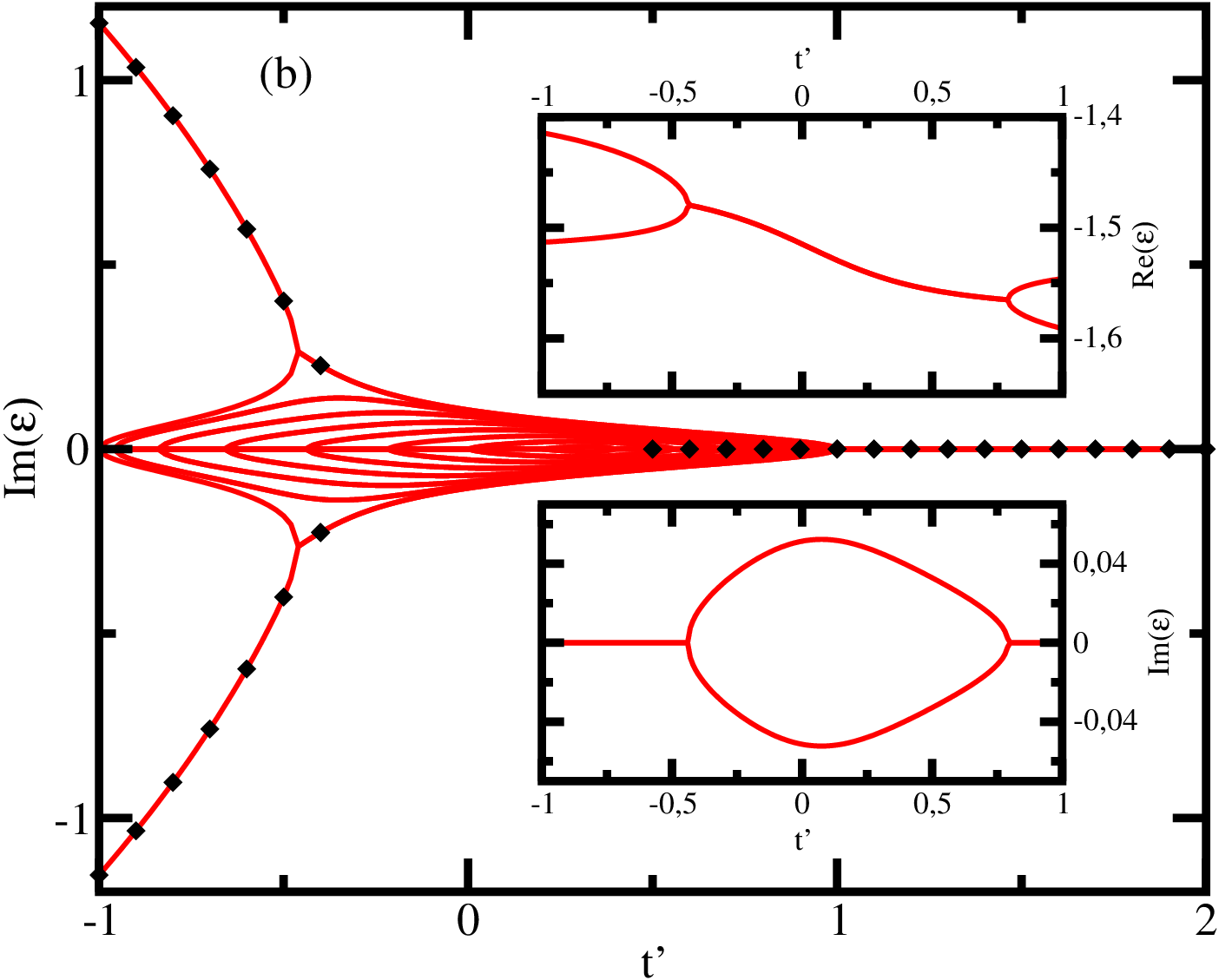}
 \caption{Energy eigenvalues of a system with Hermitian bonds with hopping $t=-1$, and one non-Hermitian impurity with left hopping $t''=3$ and right-hopping $t'$.  Panel (a) real part of energy eigenvalues.  The red lines indicates the real part of the energy eigenvalues as a function of $t'$.  The solid diamonds indicate the states and the regions for which the bulk-boundary indicator, $D_L(\epsilon)$, diverges.  Panel (b) imaginary part of the eigenvalues.  The insets of Panel (b) show two states, $7$th and $8$th in real part of the energy eigenvalue.  The upper panel shows the real part of these two energy eigenvalues, the lower the imaginary part.  The imaginary parts are finite only in regions in which the states are degenerate.}
 \label{fig:HN2BondImp_ev}
\end{figure}

Fig. \ref{fig:HNBondImp_D} shows $D_L(\epsilon)$ for the tight-binding model with a Hatano-Nelson impurity.   Fig. \ref{fig:HNBondImp_D}(a) shows a sweep of $t'$ for all states for three system sizes ($L=32, 64, 128$), the $y$-axis shown on a log scale.  At $t'=1$, for each system size, one set of results starts to increase as $t'$ is increased further.  This set is actually two states (they correspond to exactly the same number for $D_L(\epsilon)$, so it appears as one result for each system size on the graph).  The increase in these $D_L(\epsilon)$ values with $t'$ becomes steeper as the system size is increased.   From detailed analysis we find that these states are the ones which acquire an imaginary energy eigenvalue part upon crossing the point $t'=1$.  Fig. \ref{fig:HNBondImp_D}(b) shows $D_L(\epsilon)$ as a function of the real part of $\epsilon$ for a single system size $L=128$ two different $t'$ values in the two distinct regions identified in Figs.  \ref{fig:HNBondImp_ev} and  \ref{fig:HNBondImp_D}(a).  The important point is that there is an edge localized state, with a diverging $D_L(\epsilon)$, but, in contrast to the case with the Hermitian bond impurity, the real part of the energy is now zero. \\  

Fig. \ref{fig:HNBondImp_z} shows the $z$ values calculated for the model for a system with $L=32$ and the probability distribution of the right eigenvector for states mentioned above.  For $t'=0.5$ we find that all $z$ values fall on the unit circle (Brillouin zone), and the zero energy state (both real and imaginary part of the energy eigenvalue) indicates delocalization.  For $t'=1.8$ the $z$ values are mostly on the unit circle, except for four values (two states), indicated by red asterisks.   These are the $z$ values associated with the two states which have energy eigenvalues with non-zero imaginary parts.  They all fall on the imaginary axis, but not on the Brillouin zone.  They come in reciprocal and complex conjugate and also reciprocal pairs.  Panels (c) and (d) of Fig. \ref{fig:HNBondImp_z} show the extent to which certain states are localized on either side of the transition.  For $t'=0.5$ we show the probability distribution of a state with zero energy eigenvalue (both real and imaginary parts).  This state is not localized at the impurity, it is an extended state.  The state shown in panel (d) ($t'=1.8$), a state with finite imaginary part of the energy eigenvalue, shows uneven localization near the impurity, as expected for a non-Hermitian impurity. \\

\subsection{Non-Hermitian bond impurity, $t \neq t' \neq t''$}

\label{ssec:r3}

In this subsection we present the results of a scan in $t'$ which interpolates between the results of subsection \ref{ssec:r1} and \ref{ssec:r2}.  In particular, we fix $t =-1$, $t''=-3$ and we scan $t'$ in between, $t<t'<t''$. \\

\begin{figure}[ht]
 \centering
 \includegraphics[width=8cm,keepaspectratio=true]{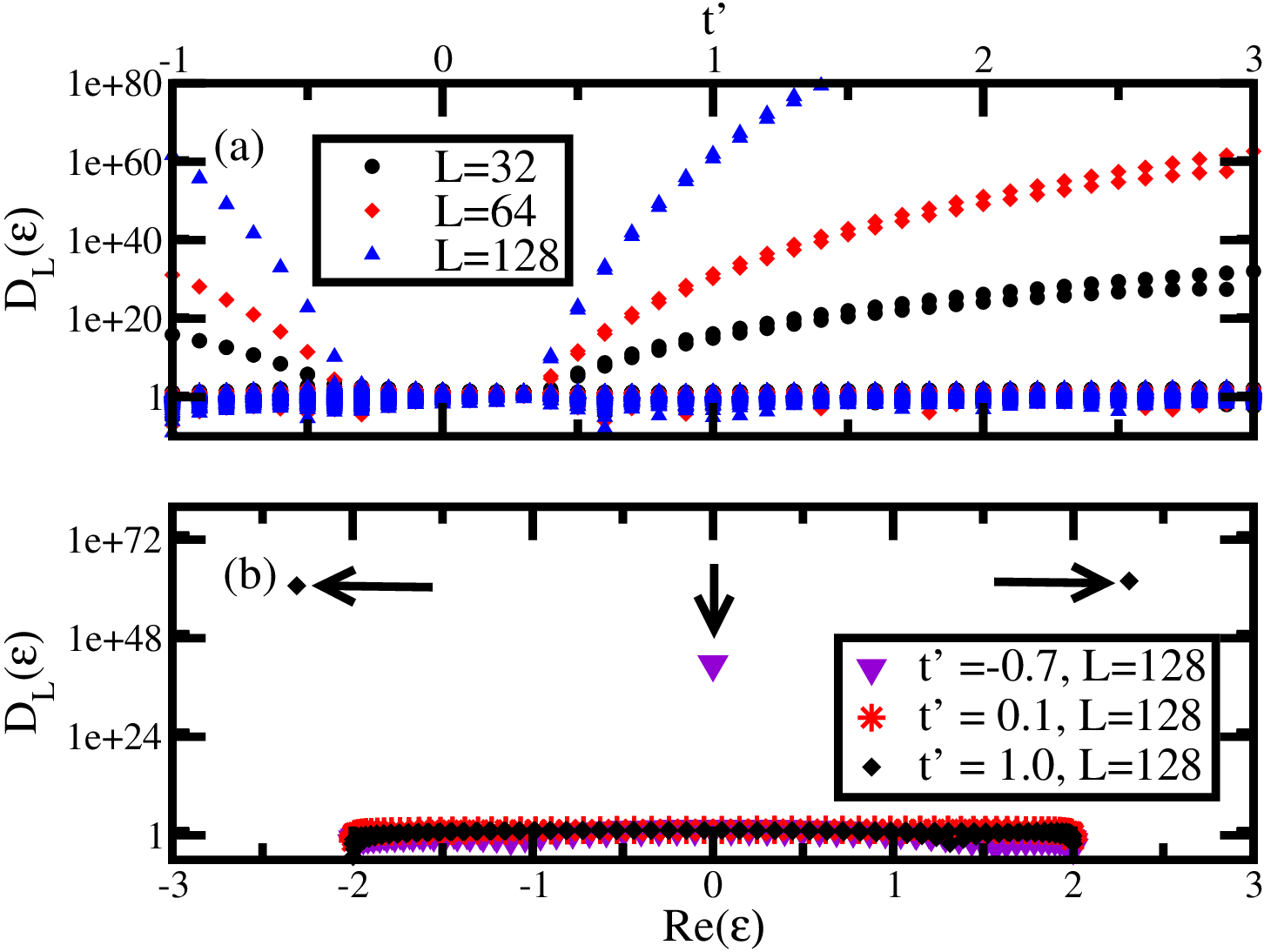}
 \caption{Bulk-boundary correspondence indicator, $D_L(\epsilon)$ for a system with $t=-1$, $t''=3$ as a function of $t'$.   Panel (a) shows $D_L(\epsilon)$ as a function of $t'$ for all states for three system sizes ($L=32, 64, 128$).  There are two regions in which diverging $D_L(\epsilon)$ are found.  Panel (b) shows $D_L(\epsilon)$ as a function of the real part of the energy eigenvalue, for three values of $t'$, $t'=-0.7,0.1,1.0$.  The arrows indicate states at which $D_L(\epsilon)$ diverges. }
 \label{fig:HN2BondImp_D}
\end{figure}

Fig. \ref{fig:HN2BondImp_ev} shows the energy eigenvalues for this scan ($L=32$), the real parts in panel (a), the imaginary parts, panel (b).  Red solid lines indicate all states, the black diamonds indicate states and regions of $t'$ where $D_L(\epsilon)$ diveges (shown in Fig. \ref{fig:HN2BondImp_D}, discussed below).  We see that as the scan proceeds, the real parts of pairs of energy eigenvalues become degenerate, and this is accompanied by imaginary parts becoming finite.  (The two insets in the figure show a zoom of this state of affairs for two states in particular, the $7$th and $8$th eigenstates of the $L=32$ system).  We see that for small $t'$ there are two states with widely diverging imaginary eigenvalues, one in the poslitive, one in the negative direction, with diverging $D_L(\epsilon)$.  These are localized states of the type found in subsection \ref{ssec:r2}.  At $t'=-1$ we see that only the localized states exhibit finite imaginary parts, which is exactly the parameter range studied in subsection \ref{ssec:r2}.  The region in which these types of localized states exist is $t' <-0.6$ in the figure.  When $t'$ is close to $t''=3$ the situation is similar to the case analyzed in subsection \ref{ssec:r1}.  There are impurity localized states, whose energy eigenvalues have no imaginary parts, and their real parts split off from the rest of the real part of the band, one in the positive, one in the negative direction.  For even system sizes we also find an exceptional point at zero energy when $t'=t$.  At this value of $t'$ the eigenvectors of the $k=0$ and $k=\pi$ states coalesce. \\

The existence of the three qualitatively different regions can also be seen from the analysis of $D_L(\epsilon)$ shown in Fig. \ref{fig:HN2BondImp_D}.  Part (a) of the figure shows $D_L(\epsilon)$ for all states scanned in the variable $t'$ for three different system sizes ($L=32,64,128$), and the two regions with localized states are clearly identified from the behavior of this quantity.  The $y$-axis is shown on a logarithmic scale, and $D_L(\epsilon)$ grows many orders of magnitude as these regions are entered for the states localized.  Part (b) of the figure shows three sets of results for a large system size ($L=128$), one in each region ($t'=-0.7,0.1,1.0$) as a function of the real part of the energy eigenvalue.  For the $t'=-0.7$ set, we see two edge states, separated from the upper and lower edges of the energy band, while for $t'=1.0$ we see a diverging $D_L(\epsilon)$ at zero (which corresponds to two localized states with zero real part of the energy eigenvalue).  The localized states are indicated with arrows.  All other states exihibit non-divergent $D_L(\epsilon)$.  Also, for $t'$ in the intermediate region ($t'=0.1$) there are no diverging $D_L(\epsilon)$ values.   \\

\begin{figure}[t]
 \centering
 \includegraphics[width=8cm,keepaspectratio=true]{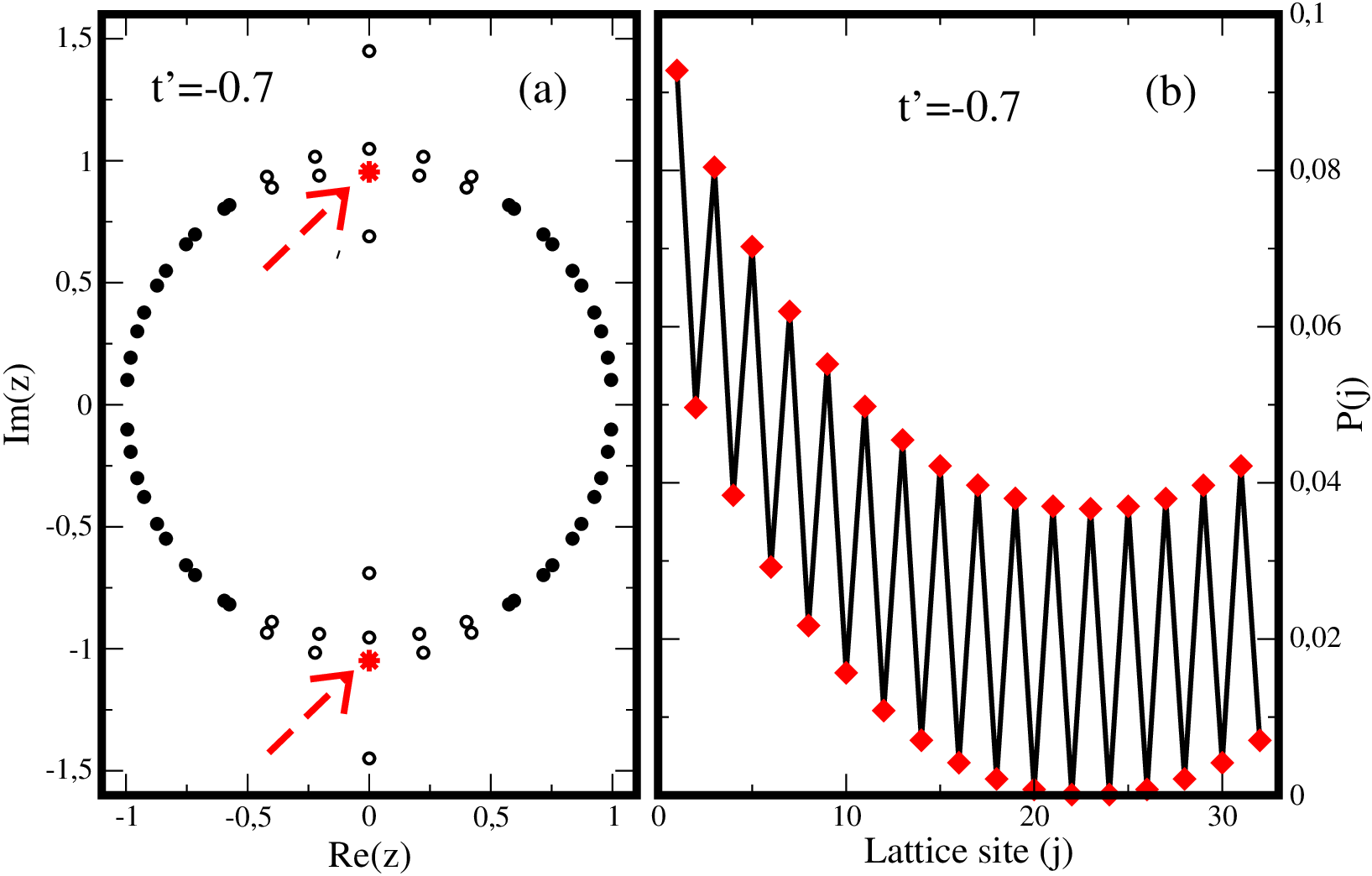}
 \includegraphics[width=8cm,keepaspectratio=true]{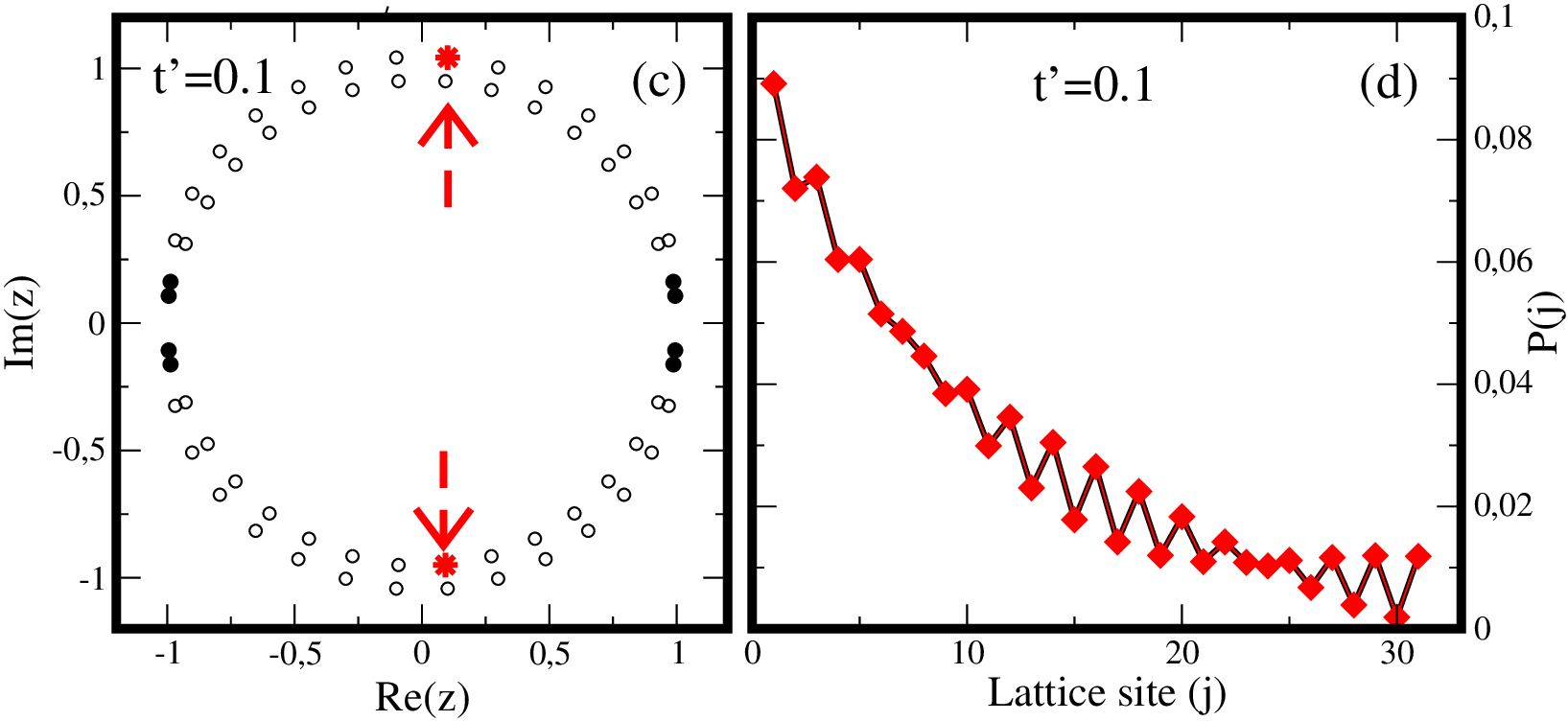}
 \includegraphics[width=8cm,keepaspectratio=true]{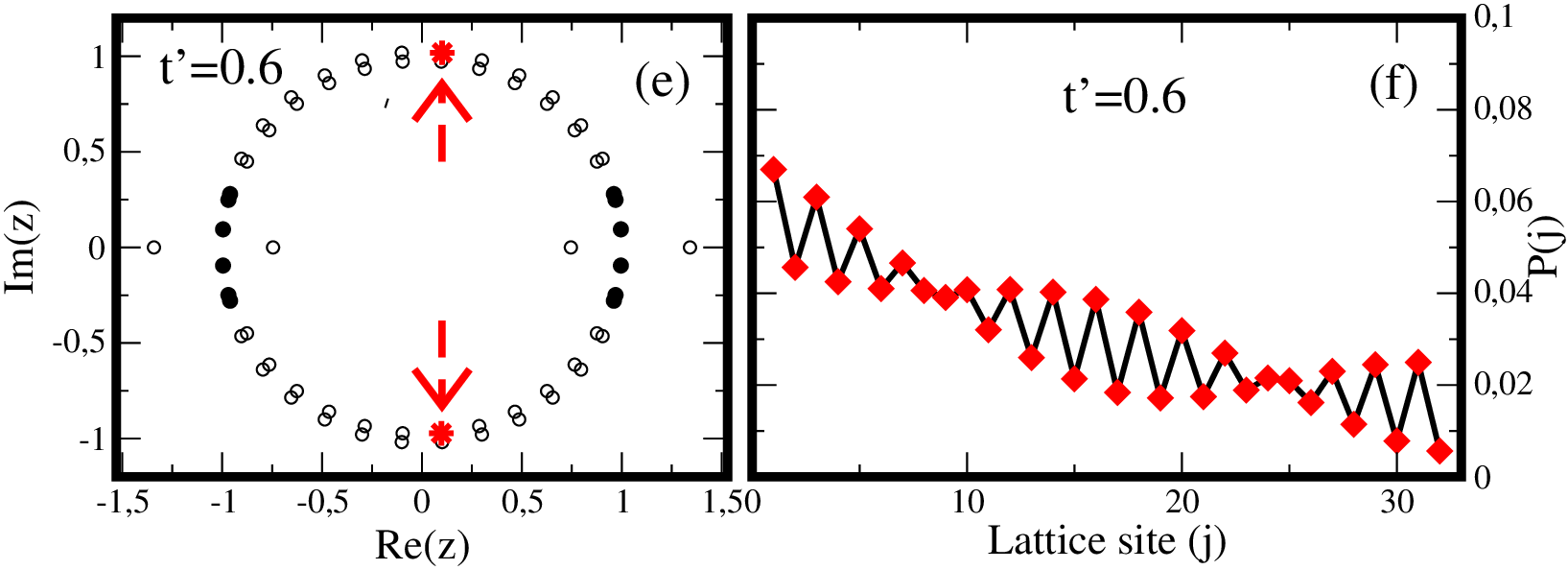}
 \caption{Panels (a), (c), (e) show the eigenvalues of the translation operator on the complex plane for a system with $t=-1$, $t''=3$, and $t'$ taking the values indicated ($t'=-0.7,0.1,0.6$.  Panels (b), (d), (f) show the probabily distribution of a chosen right eigenvector, whose $z$ values are indicated in the panels (a), (c), (e) with red dashed arrows.  In panels (a), (c), and (e) black filled circles indicate the $z$ values of states which are on the unit cirlce to within a numerical tolerance, while states with open circles are off the unit circle.}
 \label{fig:HN2BondImp_z}
\end{figure}

Fig. \ref{fig:HN2BondImp_z} shows the $z$ values on the complex plane for the three cases, $t'=-0.7, 0.1, 0.6$ and one example of the probably distribution associated with a particular right eigenvector, whose $z$ values are indicated in red asterisks, and red dashed arrows.  As the energy eigenvalues acquire imaginary parts, the $z$ values come off the unit circle, meaning that the states are no longer extended states, and localization becomes possible.   Panels (b), (d), and (f) show the probability distribution obtained from three right eigenvectors (modulus squared).  All three indicate localization, however, these states do not lead to a diverging $D_L(\epsilon)$, and in this sense they belong to a different category from the localized states find in subsections \ref{ssec:r1} and \ref{ssec:r2}.   Our further calculations indicate that in the intermediate region, a fraction of states always move off the unit circle, and exhibit localization near the impurity, in a manner similar to the skin effect for open boundaries. \\

\begin{figure}[t]
 \centering
 \includegraphics[width=8cm,keepaspectratio=true]{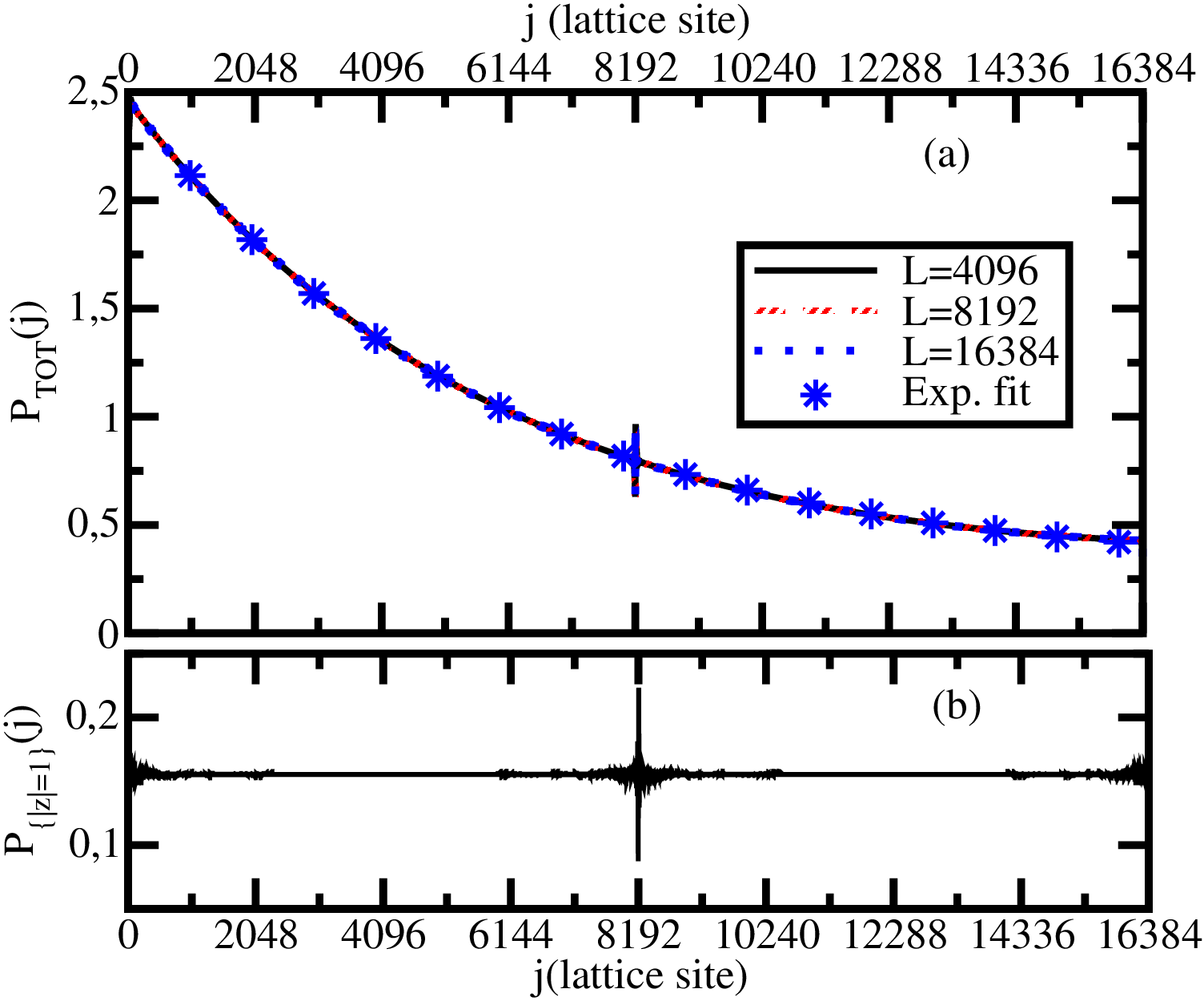}
 \caption{Panel (a) probability distribution (complex conjugate square of the wavefunction) summed over all states for a system with $t=-1$, $t'=3$ and $t''=0.1$ for three system sizes, $L=4096, 8192, 16394$.  The $j$ values of $L=4096$ system are multiplied by four, for the $L=8192$ system they are multiplied by two, to show that these functions are identical.  The blue stars indicate an exponential fit, $f(x) = a \exp(-b x) + c$ ($a = 2.17455, b = 0.00017789, c = 0.29442$).  Panel (b), probability distribution as a function of lattice site summed over states with $z$ values obey $|z|=1$ to within a numerical tolerance. }
 \label{fig:skin}
\end{figure}

Fig. \ref{fig:skin} shows sums of the complex conjugate squared wavefunctions as a function of lattice site for a system with $t=-1$, $t'=3$, and $t''=0.1$.  Panel (a) shows the sum over all states for three system sizes, $L=4096,8192,16384$.  For the first two system sizes, the $x$-axis is scaled: for $L=4096$ the scale factor is four, for $L=8192$, the scale factor is two.  This was done to show that these scaled functions are identical.  The blue stars indicate a fit of the function $f(x) = a \exp(-b x) + c$ ($a = 2.17455, b = 0.00017789, c = 0.29442$).  The results show localization around the impurity, and the scaling behavior with the lattice site indicates extensivity of the localized functions.  The localization length associated with the exponential function, $\xi = 1/b$ scales linearly with system size.  Panel (b) shows the sum of complex conjugate squared wave functions summed over those states for which $|z|=1$.  These functions are extended states, showing no localization, they are essentially flat, apart from oscillatory behavior seen at the edges and in the middle.  For further corroboration of the extensive nature of the skin effect in this model we also calculate the fraction of states for which $|z| \neq 1$ to within a numerical tolerance.  For the system sizes, $L=4096, 8192, 16384$, we find the values: $0.8447$ for all three cases.  This shows that the number of states which move off the Brillouin zone and participate in the non-Hermitian skin effect is extensive.  \\

\begin{figure}[t]
 \centering
 \includegraphics[width=8cm,keepaspectratio=true]{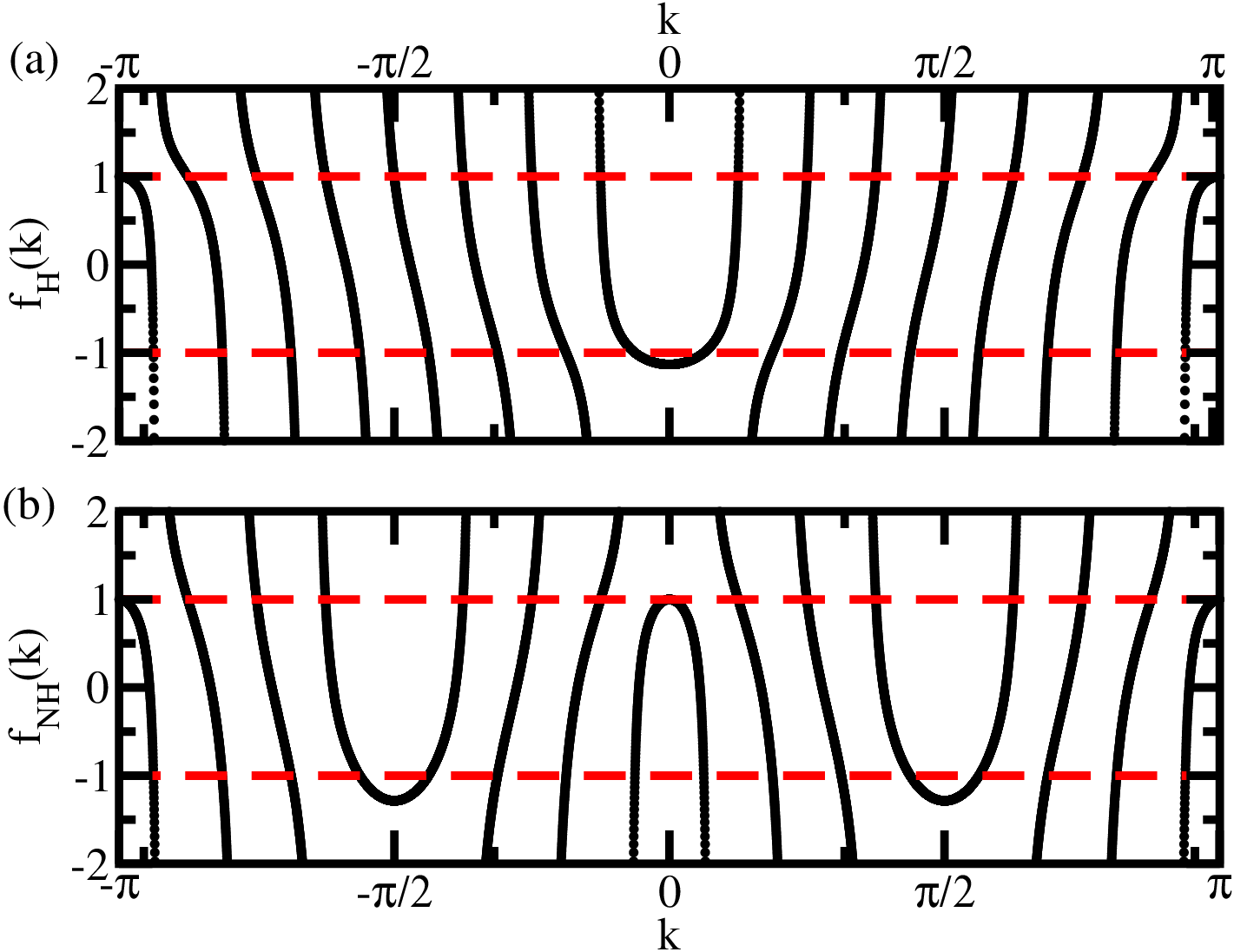}
 \caption{Graphical representation of the boundary/impurity equation for extended states ($z=e^{ik}$).  The system size is $L=16$.  Panel (a) shows the tight-binding model ($t$) with a Hermitian impurity ($t'$).  The black lines show the left-hand side of Eq. (\ref{eqn:bie_H_ext}).  Panel (b) shows the tigh-binding model ($t$) with a single non-Hermitian impurity ($t$, $t'$).  The black lines show the left-hand side of Eq. (\ref{eqn:bie_NH_ext}).  The red dashed lines in both figures correspond to $\pm1$.}
\label{fig:BIE}
\end{figure}

\begin{figure}[t]
 \centering
 \includegraphics[width=8cm,keepaspectratio=true]{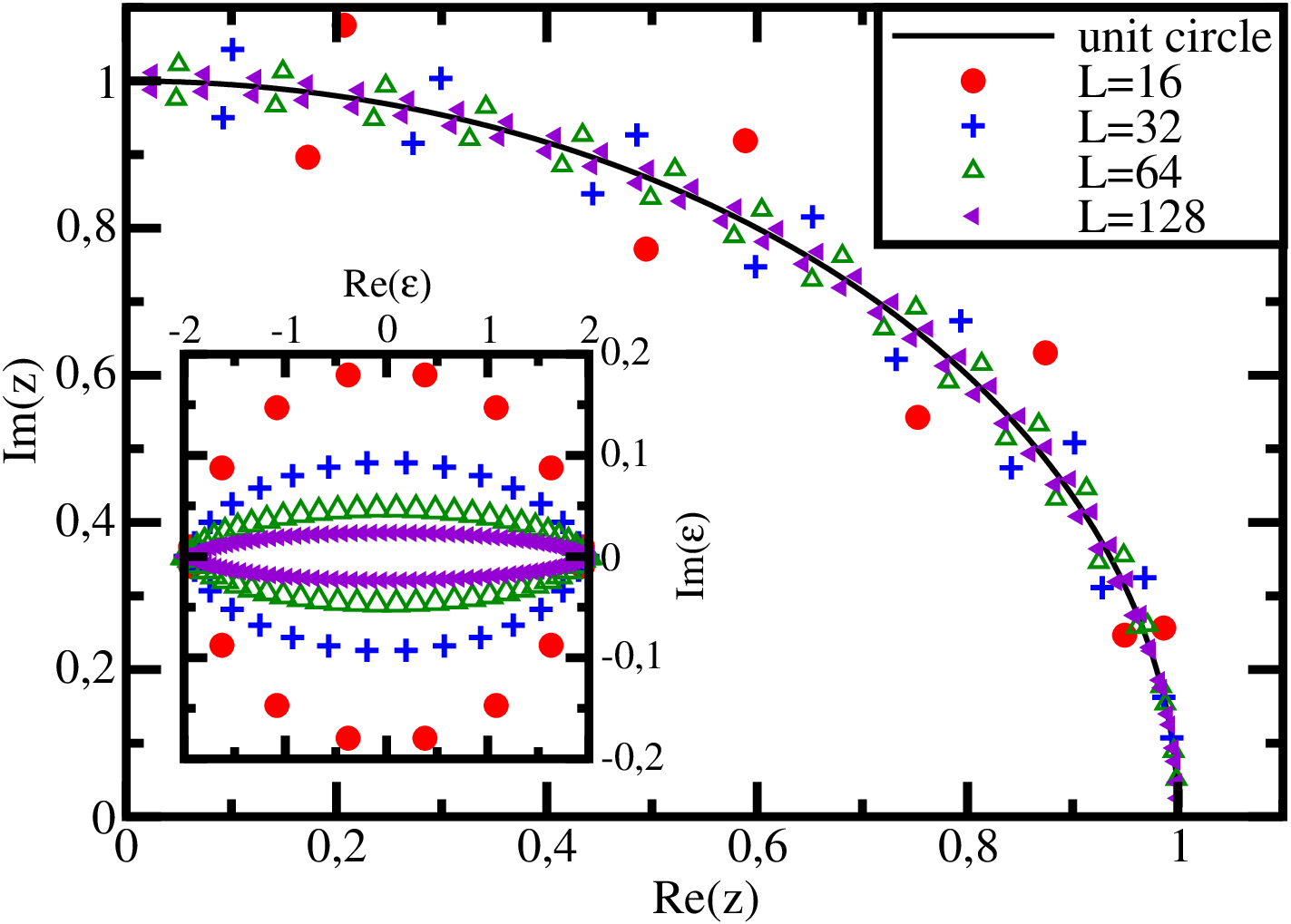}
 \caption{Eigenvalues of the translation operator ($z$-values) for different system sizes for a tight-binding model ($t=-1$) and one non-Hermitian impurity ($t'=0.1,t''=3$) shown for one quadrant of the $z$-plane.  The solid line indicates the unit circle.  As the system size increases, there are more $z$ values, and they approach the unit circle.  The inset shows the energy eigenvalues.  As the system size increases, the imaginary part is suppressed.}
\label{fig:zTL}
\end{figure}

\section{Analysis}

\label{sec:nlss}

In this section, we analyze our results using the BIE (Eq. \ref{eqn:bie}).  We show that all our results follow from this equation.  In addition, we also elucidate the role of symmetry operators in the case of edge states.

\subsection{Hermitian bond-impurity, $t'=t''\neq t$}

In the case of a Hermitian impurity ($t''=t'$), the BIE takes the form, 
\begin{equation}
\label{eqn:bie_H}
\frac{z^{L+1}\pm 1}{z\pm z^L} = \frac{t'}{t}.
\end{equation}
Restricting to extended states solutions, $z=e^{ik}$ results in the equation for the $k$-values,
\begin{equation}
\label{eqn:bie_H_ext}
f_H(k) = \frac{\cos(kL)\pm \cos(k)}{1 \pm \cos(k(L-1))} = \frac{t'}{t}.
\end{equation}

We also consider a state with $z=A$, where $A \in \mathbb{R}$.  In this case, the BIE becomes,
\begin{equation}
\frac{A^{(L+1)} \pm 1}{A \pm A^{L}} = \frac{t'}{t}.
\end{equation}
In one case, $|A|>1$.  Taking the thermodynamic limit results in,
\begin{equation}
A = \pm \frac{t'}{t},
\end{equation}
which, due to the restriction on $A$, only has a solution, if $|t'|>|t|$.  This also holds for the case $|A|<1$, where the thermodynamic limit leads to
\begin{equation}
\frac{1}{A} = \pm \frac{t'}{t}.
\end{equation}
The $+$ equations provide a solution for $\kappa$ for the case when $t'$ and $t$ both have the same sign, while the $-$ equations achieve that when the two hoppings have opposite signs.  There are no other types of  solutions to the BIE in the Hermitian case, $z$ can not be off the unit circle and have an imaginary component.  This can also be inferred from the symmetry analysis in Section \ref{sec:TBImp}.\\

Fig. \ref{fig:BIE}(a) shows a graphical representation of Eq. (\ref{eqn:bie_H_ext}) (the function $f_H(k)$ as a function of $k$) for a system of size $L=16$.   The graph shows the case for which $f_H(k)$ has a plus sign in Eq.  (\ref{eqn:bie_H_ext}).  If $t'/t$ is set equal to a value in the range $-1<t'/t<1$, then there are sixteen solutions, all states are extended.  Going outside of this range leads to the disappearance of two extended states.   These occur either at $k=0$ or $k=\pi$, both of which correspond to $z$ values on the real axis, as shown earlier in Fig. \ref{fig:HBondImp_z}. \\
\subsection{Non-Hermitian bond-impurity, $t=t''\neq t'$}

In this case, the BIE takes the form,
\begin{equation}
\label{eqn:bie_NH}
\frac{z^{L+2}+1}{z^2+z^L} = \frac{t'}{t}.
\end{equation}
Assuming $z=e^{ik}$ for extended states, one obtains an equation for the possible $k$-values,
\begin{equation}
\label{eqn:bie_NH_ext}
f_{NH}(k) = \frac{\cos(kL)+\cos(2k)}{1+\cos(k(L-2))} = \frac{t'}{t}.
\end{equation}

We look for a solution of the form $z=A$, with $A \in \mathbb{C}$.   Assuming $|A|>1$, in the thermodynamic limit, we find,
\begin{equation}
A^2 =  \frac{t'}{t}.
\end{equation}
For the case $|A|<1$, again taking the thermodynamic limit, results in,
\begin{equation}
\frac{1}{A^2} = \frac{t'}{t}.
\end{equation}
What is interesting is that if the signs of $t'$ and $t$ are different, $A$ is purely imaginary, which is what we find (see Fig. \ref{fig:HNBondImp_z}).\\

Fig. \ref{fig:BIE}(b) shows a graphical representation of Eq. (\ref{eqn:bie_NH_ext}) (the function $f_{NH}(k)$ as a function of $k$) for a system of size $L=16$.   If $t'/t$ is set equal to a value in the range $-1<t'/t<1$, then there are sixteen solutions, all states are extended.  Going outside of this range leads to the disappearance of two extended states.   In contrast to the Hermitian impurity case, here there are two points of coalescence for $t'/t=-1$ which occur at $k=\pm \frac{\pi}{2}$.  These lead to $z$ values off the unit circle and on the imaginary axis.  Fig. \ref{fig:BIE}(b) also shows coalescence points at $k=0$ and $k=\pi$, which lead to $z$ values off the Brillouin zone but on the real axis.  We did not present results for this case, but this occurs if $t'>t$ and $t$ and $t'$ have the same sign.  We remark that the situation is the opposite if the system size is even but not divisible by four.\\

\subsection{Non-Hermitian bond-impurity, $t'\neq t''\neq t$}

For this case, the important question is whether in the thermodynamic limit the $z$ values move onto the unit circle or not?  If they all did, then there would be no non-Hermitian skin effect, all states would be extended.   In Fig. \ref{fig:zTL} the $z$-values are shown in the first quadrant of the complex plane for different system sizes.  We see that as the system size increases, the $z$ values approach the unit circle, at the same time, there is always more of them, in proportion to the growing system size.\\

To answer the question, we rearrange the BIE as,
\begin{equation}
t^2 (z^{2L+2}-1) + t(t' + t'')z^L(z^2-1)+ t' t'' (z^2-z^{2L})) = 0.
\end{equation}
We restrict $z=e^{ik}$ (with $k$ a real number).  Using this substitution, and the fact the $L$ is an integer, it can be shown that the BIE is satisfied if and only if
\begin{equation}
k = 0,\pi.
\end{equation}
Other solutions can not be on the unit circle when $L \rightarrow \infty$.  In fact, Figs. \ref{fig:HN2BondImp_z} and  \ref{fig:zTL} we see that the $z$ values approach the unit circle near $k=0,\pi$.  This result suggests that the non-Hermitian skin effect shown in Fig. \ref{fig:skin} persists in the thermodynamic limit.  The inset in the same figure shows that, although the energy eigenvalues are complex at finite system sizes, they tend to fully real values in the thermodynamic limit.\\

\section{Conclusion and Outlook}

\label{sec:cnclsn}

We investigated a tight-binding model with a Hermitian and a non-Hermitian bond impurity.  We found that such a system exhibits two different types of localization around the impurity.   In the Hermitian case, and if one bond is made non-Hermitian by changing only one of the hoppings, localized edge states appear in pairs, regardless of system size.  Our study in which we extrapolated between these two cases (non-Hermitian bond impurity throughout) we found that an intermediate region exists in which localization around the impurity arises, but it is an extensive number of states responsible: a non-Hermitian skin effect.  We implemented a generalized Bloch formalism which turned out to be an efficient way to distinguish the two cases. \\

We placed emphasis on an interesting parallel between two methodologies, one developed in the context of Hermitian topological systems with a specific aim to understand the behavior of edge states~\cite{Alase16,Alase17,Cobanera17}, the other in the context of non-Hermitian systems with the aim of generalizing the concept of the Brillouin zone~\cite{Yao18,Yokomizo19,Yang20}.  Both methods are based on explicit analysis of the lattice shift operator in the two types of systems and share other significant parallels.  In our work, we applied the bulk-boundary correspondence indicator, developed in the former, to the physical systems which form the target of the latter.   In addition, we relied on the generalized Brillouin zone for our interpretations.\\

As mentioned earlier, non-Hermitian systems have been suggested~\cite{Weidemann20,Budich20,McDonald18,Konye24,Brunelli23,Wanjura20,Xue21} for a number of technological applications.  At the heart of these are the non-Hermitian skin effect and localized edge states of topological origin.   We found that both these effects are definitely present in the impurity models we studied and that varying the impurity parameters provides control over the extent these effects manifest.  To establish technological usefulness further studies would be of interest, such as the anisotropic coupling of two tight-binding chains with oppositely oriented non-Hermitian bond impurities for topological light funneling, or the response to perturbations, including time-dependence, for sensing.\\

\section*{Acknowledgments}

We thank J\'anos Asb\'oth and Andr\'as P\'alyi for helpful discussions.  This work was supported by the HUN-REN Hungarian Research Network through the Supported Research Groups Programme, HUN-REN-BME-BCE Quantum Technology Research Group (TKCS-2024/34), by the National Research, Development and Innovation Fund of Hungary within the Quantum Technology National Excellence Program (Project No. 2017-1.2.1-NKP-2017-00001), by Grants No. K142179 and No. K142652, and by the BME-Nanotechnology FIKP Grant No. (BME FIKP-NAT).

\end{document}